\documentclass[acmsmall,screen]{acmart}

\usepackage{graphics}      

\usepackage{microtype}        
\usepackage{ccicons}          

\usepackage{soul} 
\renewcommand\hl[1]{#1} 

\def\plaintitle{YouTube Recommendations and Effects on Sharing Across Online Social Platforms}
\def\emptyauthor{}
\def\plainkeywords{youtube; twitter; reddit; content quality; moderation; cross-platform}

\makeatletter
\def\url@leostyle{%
  \@ifundefined{selectfont}{
    \def\UrlFont{\sf}
  }{
    \def\UrlFont{\small\bf\ttfamily}
  }}
\makeatother
\definecolor{linkColor}{RGB}{6,125,233}
\hypersetup{%
  pdftitle={\plaintitle},
  pdfauthor={\emptyauthor},
  pdfkeywords={\plainkeywords},
  pdfdisplaydoctitle=true, 
  bookmarksnumbered,
  pdfstartview={FitH},
  colorlinks,
  citecolor=black,
  filecolor=black,
  linkcolor=black,
  urlcolor=linkColor,
  breaklinks=true,
  hypertexnames=false
}

\setcopyright{acmcopyright}
\copyrightyear{2021}
\acmYear{2021}
\acmDOI{10.1145/1122445.1122456}

\acmJournal{PACMHCI}
\acmVolume{0}
\acmNumber{0}
\acmArticle{0}
\acmMonth{0}

\begin{document}

\title{\plaintitle}
\author{Cody Buntain}
\email{cbuntain@njit.edu}
\orcid{0000-0003-4797-3726}
\affiliation{%
  \institution{New Jersey Institute of Technology}
  \city{Newark}
  \state{New Jersey}
  \country{USA}
  \postcode{07102}
}
\author{Richard Bonneau}
\email{rb133@nyu.edu}
\affiliation{%
  \institution{Center for Social Media and Politics, New York University}
  \city{New York}
  \state{New York}
  \country{USA}
  \postcode{10002}
}
\author{Jonathan Nagler}
\email{jonathan.nagler@nyu.edu}
\affiliation{%
  \institution{Center for Social Media and Politics, New York University}
  \city{New York}
  \state{New York}
  \country{USA}
  \postcode{10002}
}
\author{Joshua A. Tucker}
\email{joshua.tucker@nyu.edu}
\affiliation{%
  \institution{Center for Social Media and Politics, New York University}
  \city{New York}
  \state{New York}
  \country{USA}
  \postcode{10002}
}

\renewcommand{\shortauthors}{Buntain et al.}

\begin{abstract}
In January 2019, YouTube announced its platform would exclude potentially harmful content from video recommendations while allowing such videos to remain on the platform.
{While this action is intended to reduce YouTube's role in propagating such content, continued availability of these videos via hyperlinks in other online spaces leaves an open question of whether such actions actually impact sharing of these videos in the broader information space.}
{This question is particularly important as other online platforms deploy similar suppressive actions that stop short of deletion despite limited understanding of such actions' impacts.}
{To assess this impact, we apply interrupted time series models to measure whether sharing of potentially harmful YouTube videos in Twitter and Reddit changed significantly in the eight months around YouTube's announcement.}
{We evaluate video sharing across three curated sets of anti-social content: a set of conspiracy videos that have been shown to experience reduced recommendations in YouTube, a larger set of videos posted by conspiracy-oriented channels, and a set of videos posted by alternative influence network (AIN) channels.}
{As a control, we also evaluate these effects on a dataset of videos from mainstream news channels.}
{Results show conspiracy-labeled and AIN videos that have evidence of YouTube's de-recommendation do experience a significant decreasing trend in sharing on both Twitter and Reddit.}
{At the same time, however, videos from conspiracy-oriented channels actually experience a significant increase in sharing on Reddit following YouTube's intervention, suggesting these actions may have unintended consequences in pushing less overtly harmful conspiratorial content.}
{Mainstream news sharing likewise sees increases in trend on both platforms, suggesting YouTube's suppression of particular content types has a targeted effect.}
{In summary, while this work finds evidence that reducing exposure to anti-social videos within YouTube potentially reduces sharing on other platforms, increases in the level of conspiracy-channel sharing raise concerns about how producers -- and consumers -- of harmful content are responding to YouTube's changes.}
{Transparency from YouTube and other platforms implementing similar strategies is needed to evaluate these effects further.}

\end{abstract}



\maketitle

\section{Introduction}

\hl{As of January 25, 2019, YouTube has announced an initiative to improve the quality of content on its platform and what the platform recommends to its users} \cite{YouTubeTeam2019}. 
\hl{This initiative claims to use a combination of machine learning and human evaluation to identify videos ``that could misinform users in harmful ways'' and videos that present ``borderline content'' -- a nebulous class of video YouTube declines to define in detail.}
\hl{YouTube then purportedly suppresses these videos by removing them from recommendation (e.g., they are excluded from the ``Recommended'' section on users' homepages and the ``Up Next'' scroll after videos) while allowing the videos to remain on the platform} \cite{YouTubeTeam2019}.
\hl{Such an approach may reduce exposure to harmful content \emph{and} incentivize the production of less-harmful and more informative content, as those videos will receive more exposure through recommendation.}
\hl{This use of a large-scale recommendation engine to balance information quality (i.e., suppressing ``harmful content'') and free speech (i.e., by allowing these videos to remain available), however, has significant implications for ongoing debates around YouTube's role in radicalization and platform-level censorship and for the information ecosystem at large.}
\hl{Likewise, little empirical study has evaluated the efficacy of such approaches beyond YouTube's boundaries or their downstream consequences}.
\hl{This paper sheds light on these concerns by estimating YouTube's suppressive effects on anti-social video-sharing in Twitter and Reddit before and after YouTube's announcement, a particularly important question given the academic debate around these issues.}

\hl{One such question concerns the interactions between YouTube's recommendations and the societal factors promoting the spread of potentially harmful and misinforming content.}
\hl{On the one hand, YouTube has received considerable criticism for potentially radicalizing individuals}, both in popular press (e.g., Tufekci's opinion piece in the \textit{New York Times} \cite{Tufekci2018}) and in the academy (e.g., Ribeiro et al. \shortcite{Ribeiro2019}), so efforts to address these radicalization pathways could be a social good.
\hl{On the other hand, research such as Ledwich and Zaitsev} \cite{Ledwich2020} \hl{and Munger} \cite{Munger_2019} \hl{find the opposite, arguing instead that anti-social content is popular on YouTube, not just because of recommendation engines, but also because of the political economies of YouTube's audience and content producers} \shortcite{Munger_2019,Munger2019}.
\hl{Munger and Phillips, in particular,} argue that the expanding supply of alternative (and especially politically far-right) content on YouTube is driven by a demand that is unsatisfied by traditional media sources, rendering YouTube's actions  insignificant in the face of a larger societal issue.

{At the same time, YouTube does not exist in isolation, and efforts to address its role in propagating harmful content must be evaluated across the larger information space}.
While other works have shown YouTube has reduced recommendations to conspiratorial and alternative news content \emph{within its own platform} \cite{Faddoul2020,Suzor2019}, exposure to such content occurs in many places beyond YouTube.
Instead, links to YouTube content are routinely some of the most shared across Facebook, Twitter, and Reddit, with a recent study of COVID-19 discourse across five different online social platforms finding YouTube videos shared therein are significantly amplified \cite{Cinelli2020}.
\hl{This multi-platform perspective is also essential as many social media platforms are increasingly leveraging ``de- recommendation'' for computationally supported content moderation (e.g., Facebook now suppresses articles third-party fact-checkers rate as false} \cite{Facebook2018}, \hl{and Twitter filters false content from user feeds,\footnote{\url{https://help.twitter.com/en/safety-and-security/tweet-visibility}} to name a few).}
\hl{As such, whether a single platform's unilateral actions are sufficient to influence propagation of the harmful content YouTube is targeting is unclear}.

\hl{This paper engages with these questions by answering the following: \emph{Does YouTube's action have a significant impact on sharing patterns of harmful and misinforming YouTube videos on other online social networking platforms?}}
\hl{We divide this question across two axes: the online social platform and the different types of anti-social videos.}
At the platform level, we examine whether YouTube's change has impacted sharing on Twitter and on Reddit. 
At the video level, because YouTube does not publish or identify videos it has flagged as potentially misinforming or harmful, we cannot directly examine impact on these ``potentially harmful'' videos.
Instead, we rely on related work that has been collecting in-situ recommendations for several years \cite{Faddoul2020,Suzor2019}.
\hl{From these sources, we build three datasets of anti-social content to evaluate YouTube's changes and compare them to a fourth control set of mainstream media content.}
\hl{These three sets of potentially harmful or misinforming content contain: 1) videos labeled as conspiracy, 2) videos from channels that specialize in conspiracy content, and 3) videos from politically extreme channels that have been identified as part of the ``Alternative Influence Network'' (AIN)} \cite{Lewis2018}.

To evaluate the impact of YouTube's action on these platforms and video types, we compare sharing frequencies before and after YouTube's announcement using a form of comparative interrupted time series (ITS) analysis.
\hl{ITS models have been similarly used to assess the impact of other content moderation policies, as in Reddit's banning of several anti-social communities in 2015, as described in Chandrasekharan et al.} \cite{Chandrasekharan:2017:YCS:3171581.3134666}.
\hl{These ITS models allow us to construct a study design for evaluating YouTube's changes as a population-level intervention that is 1) deployed at an approximately discrete moment in time, 2) have expected population-level outcomes, and 3) whose effects are expected to compound over time.}
\hl{In particular, these models are well-suited to retrospective studies where such interventions have been deployed but randomized, controlled trials are either unavailable or prohibitive.}
\hl{We apply these models to daily time-series data of YouTube sharing on Twitter and Reddit for a collection of video types (i.e., the treatment groups) that should be impacted by YouTube's changes and compare these time series to a control group of videos (mainstream media videos).}
\hl{In these time series, we identify the point at which YouTube announced its recommendation changes as the point where the time series is ``interrupted'' and compare the observed time series after this interruption to an expected trend generated by forecasting forward from the pre-interruption data.}
\hl{Differences between the observed and expected time series data, expressed through coefficients in a regression model, capture the intervention's effects, both in terms of a sharp, level change and a compounding change in trend.}
\hl{For a more in-depth discussion and tutorial on ITS analysis, see Bernal, Cummins, and Gasparrini} \cite{Bernal2017}.

\hl{Using these models, our analysis suggest YouTube's de-recommendation strategy has a significant suppressive effect on both conspiracy-focused videos and AIN-sourced videos, with results showing a significant, consistent, and negative trend in sharing in both Twitter and Reddit.}
\hl{These results are consistent with a compounding effect of YouTube's changes, which gradually reduces cross-platform sharing of the treated content over time (though alternative explanations exist, which we discuss later).}
Effects on mainstream media sharing are consistent with our expectation in that we see no suppressive effect on either platform; rather, we actually find a significant \emph{increase} in mainstream media sharing, in trend on both platforms and in level on Reddit.
\hl{For videos produced by conspiracy-oriented channels, however, we find Reddit experiences a significant increase in the level of sharing, contrary to expectation, and despite null effects on Twitter.}

\hl{Following these results, we reflect on the potential normative effects YouTube's change has on conspiracy- and AIN-video sharing while balancing these results with the increase in content from conspiracy-oriented channels.
In this reflection, we discuss the tradeoff between deleting content compared to limiting its propagation, with the recent popularity of the \emph{Plandemic} anti-vaccine video as a motivating example.
We also discuss potential explanations for the growth of conspiracy-oriented channels, as this finding points to potentially unanticipated amplification effects of content adjacent to conspiracy-labeled videos.
We also acknowledge that, while we examine only a strict subset of all possible ``potentially harmful and misinforming'' videos, numerous other types of potentially harmful videos exist (e.g., hate, racism, sexism, ``fake news'', etc.), and how these videos are impacted by YouTube's moderation efforts are open and important questions.}

\hl{Though results in this work indicate an effective approach to platform-level moderation, this study has important limitations that inhibit strong claims about effects and outcomes, and we conclude with an in-depth evaluation of potential threats to validity and related limitations in this work.
In particular, YouTube's reticence to identify videos to which the de-recommendation treatment has been applied complicates identifying treated videos and when these videos were subjected de-recommendation.
Likewise, our observations are limited to video sharing rather than viewership, so it remains possible that overall video views are unaffected by YouTube's actions (though seemingly unlikely).
While section 7 covers these threats to validity and other limitations in more detail, these results provide evidence that YouTube's actions -- and de-recommendation in general -- may be a viable approach to improving information quality.
Thoroughly addressing such threats, however, requires close collaboration with YouTube and other mainstream online social platforms.}

\subsection{Contributions and Policy Implications}

\hl{This work presents several contributions, both for research into platform-level content moderation strategies and for policy.
First, this effort provides a template for future studies of platform-level moderation practices, as the ITS models we employ are well-suited to this context.
This methodology can be applied to Facebook's suppression of false articles, Reddit's quarantined communities, Twitter's suppression of inaccurate tweets, and other similar cases.
At the same time, while this work observes reductions in anti-social YouTube sharing on other platforms, increases in the level of conspiracy-channel sharing raise concerns about how producers -- and consumers -- of harmful content are responding to YouTube's changes.
These findings open new questions about the potential unintended side-effects of this approach -- in particular, identifying what content receives the redirected recommendations is crucial for understanding impact on information quality.
Otherwise, platforms run the risk of suppressing a particular kind of anti-social content only to unknowingly preference a different anti-social behavior.}

\hl{Policy-wise, our work suggests that removing potentially harmful content from recommendation could strike an effective balance whereby free-speech considerations are respected without actively propagating anti-social or subversive content. 
As outright deletion of undesirable content can feed into anti-establishment narratives, thereby unintentionally accelerating the spread of anti-social content, de-recommendation may also be a viable approach to counter these frames.
Lack of transparency around these interventions, however, risk amplifying concerns about platform censorship while also inhibiting research into these strategies.
New methods for providing transparency from YouTube and other platforms implementing similar strategies are therefore needed to address these concerns and evaluate these effects further. }


\section{Background}

\hl{YouTube's effort is one part of a larger effort by Google, Facebook, and other companies to address radical and extremist content on their platforms.}
Such attempts have been controversial as opponents to these efforts claim the platforms are censoring their users and violating free speech.
A point stressed in YouTube's announcement, potentially in response to these claims, is that their change ``will only affect recommendations of what videos to watch, not whether a video is available on YouTube'' \cite{YouTubeTeam2019}.
This move can then be viewed as a compromise, still allowing potentially fringe ideas a platform but without actively spreading these ideas through recommendations.
Less diplomatic efforts, however, have had normative effects: Reddit went through a similar cycle of controversy with their outright banning of several hateful communities in 2015, and while contentious, research has since shown this ban had a socially normative effect on discourse within Reddit \cite{Chandrasekharan:2017:YCS:3171581.3134666}.

YouTube's compromise in allowing these potentially misinforming/harmful videos to remain available has important implications since YouTube is not an isolated platform; rather, it exists within a broader information ecosystem. 
This ecosystem is comprised of numerous online information sources, from video platforms (e.g., YouTube and Twitch) to social media platforms (e.g., Facebook, Instagram, Twitter, and Reddit) to point-to-point messaging apps (e.g., WhatsApp or WeChat) to online news websites (e.g., the New York Times or Wall Street Journal's websites or Internet-native news like BuzzFeed) and blogs (e.g., Tumblr, WordPress, etc.).
As the median American uses three separate social media platforms \cite{Smith2018}, limiting exposure on a single platform (even one as large as YouTube) may have little effect on this content's uptake, especially if the primary avenue for individuals to find this content is via other platforms.
If a conspiracy-focused community on Reddit, Twitter, or Facebook is creating and sharing conspiracy-focused YouTube videos in these other spaces, such videos may still propagate rapidly across the information ecosystem regardless of whether they are being recommended natively within YouTube. 

The power of these fringe communities in sharing anti-social content is already well-supported in the literature \cite{Zannettou:2017:WCU:3131365.3131390,Mariconti2019}, showing ``tight-knit communities'' in Reddit and other platforms spread content across the information ecosystem.
Since YouTube is consistently one of the most popular platforms globally and one of the most popular domains linked to in Twitter and Reddit, the spread of these videos therefore may not rely on YouTube at all.
Individuals' primary vector for exposure to this content may instead be through interactions with other conspiracy-minded people rather than YouTube's on-site recommendation system.
Alternatively, YouTube's recommendation algorithm may be the primary vector through which otherwise unexposed individuals are shown this content (though academic consensus on this issue has not been reached, e.g.,  \cite{Ribeiro2019,Tufekci2018,Ledwich2020,Munger_2019}), and protecting individuals natively within YouTube could be an effective means to reduce exposure.

YouTube's efforts also appear to be more than lip-service as well, with other researchers finding evidence of significant in-platform reductions of recommendations to certain classes of anti-social videos.
Work by Faddoul et al. \cite{Faddoul2020} in particular demonstrates YouTube's recommendation system has reduced recommendations to conspiratorial content within the platform.
That work has developed a set of 6,752 conspiracy-focused videos by collecting the daily top 1,000 most recommended videos from a seed set of 1,080 information- and news-oriented channels between October 2018 and February 2020.
From these daily collections, each of the resulting 8 million recommended videos is then passed through a binary classifier for detecting conspiratorial content, which Faddoul et al. have trained on the video's transcript, ``snippet'' content (i.e., title, description, and tags), and comments and validated against a set of videos collected from conspiracy-oriented communities in Reddit.
Their classifier achieves a precision of 0.78 and recall of 0.86 \cite{Faddoul2020}, suggesting the majority of the conspiracy-labeled videos are correctly identified as conspiracy-oriented, and the large majority of conspiracy-oriented videos are captured in this set.
Using these conspiracy videos, Faddoul et al. perform a longitudinal analysis of YouTube's recommendations and show a marked decrease in recommendation of these conspiracy videos in April 2019, with a minimum in May 2019.
These results provide strong evidence that this set of videos has been subject to de-recommendation within YouTube's platform.

A technical report from Digital Social Contract, a program housed at Queensland University of Technology's Digital Media Research Centre, describes a similar longitudinal study of YouTube recommendations but performed on videos sourced from AIN channels rather than conspiracy-specific channels \cite{Suzor2019}.
That report shows, of in-platform recommendations from a sample of 3.6 million YouTube videos, videos posted by AIN channels were recommended 7.8\% of the time in the first two weeks of February.
In the two weeks after February 15, however, these recommendations have dropped substantially to 0.4\% of recommendations, suggesting AIN-authored videos have seen a steep decline following YouTube's announcement.

While these works provide substantial evidence that YouTube's change to its recommendation systems does appear to impact \emph{within-platform} recommendations, we are still left with a critical question: \emph{Does YouTube's choice to modify its internal recommendations without actually deleting this content significantly impact the propagation of this content across the information ecosystem?} 
If so, the financial incentive to produce this content may be reduced through smaller audiences, but if not, YouTube may need to take bolder steps to mitigate its role in the propagation of harmful content online.
This question also has broader implications in how we should craft solutions for improving the information space: Are the actions of a single entity sufficient in this highly interconnected environment, or must the platforms work together to address these concerns? 
YouTube's decision to stop recommending this content gives us a unique opportunity to explore this interconnection, and this paper presents a {novel} analysis of these interactions across millions of social media posts and videos.

\subsection{Prior Work in Online Moderation}

%
%
%
%
%

Moderation in online social platforms is a well-studied problem, with most research existing on a spectrum between user perceptions of the moderation process (e.g., Myers West \shortcite{MyersWest2018}) to analyses of the moderation strategy's efficacy (e.g., Chandrasekharan et al. \shortcite{Chandrasekharan:2017:YCS:3171581.3134666} and Chancellor et al. \cite{Chancellor:2016:TIC:2818048.2819963}), with hybrid approaches in between (e.g., Newell et al. \cite{Newell2016}).
This paper fits in the latter part of this spectrum, specifically exploring how YouTube's announcement and ensuing action have impacted the spread of potentially harmful content beyond YouTube's boundaries. 
The majority of similar work that studies the impact of moderation does so by identifying when a new moderation policy went into effect on a given platform and evaluating changes in user behavior or sharing within that same platform.
While that approach has led to important findings, such as banning communities in Reddit increasing overall platform quality \cite{Chandrasekharan:2017:YCS:3171581.3134666}, or that users often find ways around content-based moderation strategies \cite{Chancellor:2016:TIC:2818048.2819963}, it is difficult to apply to content creation platforms that do not directly support in-platform sharing, like YouTube. 
YouTube sharing, on the other hand, is intrinsically cross-platform, with the YouTube sharing button providing options for sharing to other platforms, like Twitter, Reddit, Facebook, etc.
As a result, applying similar strategies to evaluate YouTube moderation are difficult since evaluation must necessarily use observations from other platforms.
Therefore, an important novelty of this paper is its use of cross-platform data to evaluate the impact of YouTube's moderation.

Of the prior work, Chandrasekharan et al. is most similar to our research \cite{Chandrasekharan:2017:YCS:3171581.3134666}.
In that effort, the researchers have used a similar ITS construct to evaluate changes to both discussion and user populations of communities within Reddit after Reddit banned several hateful and harassing communities.
While our work is differentiated by platform and our cross-platform context, Chandrasekharan et al. is particularly germane as one could imagine Reddit as a microcosm of the internet, with interconnected but separate online spaces analogized by Reddit's subreddit structure.
In this analogy, a change to one subreddit's moderation then indirectly impacts other subreddits through changes in population, cross-posting behavior, and language.
In fact, Reddit has taken similar steps to YouTube's de-recommendation by quarantining specific communities Reddit moderators have deemed to be in violation of their terms of service, as they did with the popular but controversial far-right subreddit \texttt{/r/the\_donald}, a step that prevents content in that community from appearing on Reddit's front page \cite{Vigdor2019}.
Likewise, Facebook and WhatsApp have taken similar steps to suppress the spread of problematic content by preventing WhatsApp users from broadcasting a link more than five times \cite{Greenspan2019}.
Both Reddit and Facebook's steps are similar to YouTube's efforts in that individuals are not banned from communicating their views, but the platforms take steps to limit exposure of those views.
As such, applying a study similar to Chandrasekharan et al. to investigate YouTube's moderation approach, as we do here, has important implications beyond YouTube.

\section{Constructing Treatment and Control Groups of YouTube Videos}

{To evaluate the effect of YouTube's de-recommendation treatment on misinforming and otherwise harmful videos, we develop a series of treatment groups that include videos likely to have been subjected to this treatment.
We likewise construct a control group, consisting of videos that should not have been subjected to this treatment, for comparison purposes.
We also acknowledge the data sources and group constructions described below have biases and limitations.
Section 7 provides an in-depth discussion of threats to validity introduced by these decisions.
}

\subsection{Identifying Potentially Harmful YouTube Videos}

We are particularly interested in whether YouTube's change affects the prevalence of anti-social content in other platforms.
To this end, we must identify ``potentially harmful or misinforming'' YouTube videos, but no video metadata appears to be available in YouTube's APIs that would suggest whether a video is being or has been removed from recommendation. 
While Google, YouTube's affiliate company, makes its guidelines for human annotation public in its ``Search Quality Evaluator Guidelines''\footnote{\url{https://static.googleusercontent.com/media/guidelines.raterhub.com/en//searchqualityevaluatorguidelines.pdf}}, and these guidelines include sections on ``Potentially Harmful Pages'' and ``Pages that Potentially Misinform Users'', it is unclear whether and how these guidelines also apply to YouTube videos.
As a result, we lack a YouTube-native method to identify videos that have been flagged for removal from recommendation, meaning it is difficult to test directly whether these same videos are shared less on other platforms. 

This exploration therefore studies three sets of videos that evidence suggests have been affected by YouTube's modifications:

\subsubsection{Conspiracy-Oriented Videos} Our first set consists of the 6,752 videos curated by Faddoul et al. \cite{Faddoul2020}, as these videos have experienced a decline in recommendation within YouTube's platform.
As this decline is potentially attributable to YouTube's internal changes, evaluating how these videos are shared on other platforms will provide important insights about cross-platform effects.
Further motivating our focus on these videos are articles in the popular press that report YouTube's efforts as targeting ``conspiracy'' videos \cite{Wakabayashi2019}.
Though YouTube has not explicitly referred to the targeted content using that label, popular discussion characterizes it as such.
Furthermore, the recent COVID-19 pandemic and resulting conspiracies has had potentially significant public-health consequences \cite{Cinelli2020,Bursztyn2020}, so evaluating cross-platform impact of such health-related information will inform decisions on how to combat health misinformation.
We refer to this set as the Conspiracy Video set.

\subsubsection{Videos from Conspiracy-Oriented Channels} {Beyond conspiracy videos, we also evaluate whether \emph{creators of conspiracy content} are affected by this change more generally, as such effects could drive more pro-social content creation.
We thus also include an expanded set of videos that have been produced by conspiracy-focused channels, which we define as those channels that are prevalent in the Faddoul et al.} \cite{Faddoul2020} dataset.
{This definition follows similar approaches by Allcott and Gentzkow} \cite{10.1257/jep.31.2.211}, {Grinberg et al. }\cite{Grinberg2019}, {Guess et al.} \cite{Guess2018}, {and others, who propagate ``fake news'' labels from individual articles to the domains that post.}
{Additionally, both Grinberg et al.} \cite{Grinberg2019} {and Starbird} \cite{Starbird2017} {have found that, in other contexts, a majority of alternative content is produced by a small set of sources, further reinforcing the notion that channels prevalent in the Faddoul et al. dataset are likely conspiracy-focused.}

{Faddoul et al. includes 6,752 videos with a conspiracy label, produced by 1,795 different channels. 
This distribution of conspiracy videos per channel is skewed, with many channels producing only a single video in the dataset (mean = 3.7616 videos per channel, median = 1, max = 203). 
The top 5\% of channels in this dataset produce at least 15 conspiracy-labeled videos per channel, which we use as cutoff to construct a list of conspiracy-focused channels, of which 80 remain active on YouTube at the time of writing.}
{Hence,} we expand our work to include 85,818 videos produced by these 80 conspiracy-oriented channels.
These channels include sources like ``Age of Truth TV'', ``Destroying the Illusion'', and ``Zohar StarGate Ancient Discoveries'' and are unified around topics like alternative histories, battling ``corporate propaganda'', and political conspiracies.
{This paper refers to this collection} as the Conspiracy Channel set.

\subsubsection{AIN Videos} Our third test set is composed of 40,764 politically extreme videos curated from a collection of ``alternative'' YouTube channels, dubbed the AIN \cite{Lewis2018}.
These AIN channels provide ``an alternative media source for viewers to obtain news and political commentary'' that ``facilitates radicalization'' \cite{Lewis2018} and therefore appear to satisfy YouTube's definition of misinforming/harmful content (though we acknowledge no publicly available mechanism exists to identify definitively what videos YouTube has labeled).\footnote{A selection of these channels have already been deleted by YouTube.} 
As videos posted to these channels experienced a steep decline in recommendation in February 2019 \cite{Suzor2019},  these videos appear to overlap with YouTube's classification scheme and provide an alternative, non-conspiratorial set of information sources when compared to the above two test sets. 
Using these AIN channels as a source, we have identified 40,764 unique videos from 69 channels using YouTube's API. 
We refer to this set as the AIN set.

\subsection{Building a Control Set of YouTube Videos}

To evaluate whether YouTube's treatment effects are specific to the video types of interest, {we construct a dataset of videos that likely \emph{are not} subject to YouTube's de-recommendation treatment}.
Such a control set should be similar in nature to the targeted content but have a low likelihood of receiving this de-recommendation treatment.
Mainstream news is a good such set in that it is similar in topic to the political content touched on in both conspiracy and AIN videos.
Likewise, as suggested within Google's own Search Quality Evaluator Guidelines, reviewers should ``find high-quality, trustworthy sources to check accuracy and the consensus of experts'', mainstream sources are considered generally high-quality and trustworthy.
As such, videos posted by mainstream news YouTube channels should be unlikely to be classified as potentially harmful or misinforming.
To this end, we have identified 65 mainstream news sources from across the political spectrum  (e.g., CNN, NPR, Fox News, Bloomberg, etc.) and their related YouTube channels.
Again, using YouTube's API, we have collected all videos published by these channels as of March 2020, resulting in 790,377 videos.
We refer to this set as the MainStream Media (MSM) set.

To ensure our control dataset is sufficiently separate from our test datasets, Table \ref{tab:video_overlap} shows the overlap among all four video types.
As the table illustrates, the mainstream news video set has very small overlap with the conspiracy set and no overlap with the other two sets.
Manual analysis of these 28 conspiracy videos that appear in the MSM set suggest some of these videos are false positives from the Faddoul et al. conspiracy classifier as the majority of videos discuss unidentified flying objects (UFOs) (e.g., ``Inside the Navy's 2015 encounter with a UFO'', ``Neil deGrasse Tyson: UFO doesn't mean aliens'', and ``New spy plane? You be the judge'').

\begin{table}[h!]
\caption{Overlap in YouTube Video Sets. Each cell represents the number of videos in common between the video set in the column and row.}
\small
\begin{center}
\begin{tabular}{p{1.7in} r r r r}

\hline
 & \multicolumn{1}{p{0.75in}}{\raggedleft\textbf{Conspiracy Videos}} & \multicolumn{1}{p{1in}}{\raggedleft\textbf{Conspiracy Channel Videos}} & \textbf{AIN Videos} & \textbf{MSM Videos} \\ \hline

\textbf{Conspiracy Videos} & 6,752 & 2,615 & 32 & 28 \\
\textbf{Conspiracy Channel Videos} & 2,615 &  85,830 & 0  & 0 \\
\textbf{AIN Videos} & 32 & 0 & 40,764 & 0 \\
\textbf{MSM Videos} & 28 & 0 & 0 & 790,163 \\

\hline

\end{tabular}
\end{center}
\label{tab:video_overlap}
\end{table}%

{Regarding the difference between the conspiracy-video and -channel datasets, 3,765 of the 6,752 conspiracy videos are authored by the 1,705 channels we excluded from our conspiracy-channel dataset.
We do note, however, that the expected overlap between these two datasets should be 2,987 videos, meaning 372 are left unaccounted.
}
When we visit a subset of these missing videos, however, we see many of them have been deleted or made private.

\begin{table}[tb]
\caption{{Mathematic Notation and Variables. For clarity, we summarize the notation and measures used in this paper.}}
\begin{center}
\small
\begin{tabular}{l p{4.5in}}

\hline
\multicolumn{1}{l}{\textbf{Variable}} & \multicolumn{1}{l}{\textbf{Definition}} \\ \hline

$p$ & Platform, either Twitter (T) or Reddit (R) \\
$t$ & Time, in days \\
$c$ & Denotes the conspiracy treatment group \\
$cc$ & Denotes the conspiracy-channel treatment group \\
$a$ & Denotes the AIN treatment group \\
$m$ & Denotes the mainstream news control group \\
$g$ & Video group, in $\{c, cc, a, m\}$ \\
$n_{t}$ & Number of posts at time $t$ \\
$v_{t}$ & Number of videos (in any group) in our dataset at time $t$ \\
$v_{g,t}$ & Number of videos in group $g$ in our dataset at time $t$ \\

$\rho_{t}$ & Proportion of posts containing YouTube videos over all posts at time $t$ \\
$\rho_{g,t}$ & Proportion of videos in group $g$ over all YouTube videos at time $t$ \\

$\overline{\rho}$ & Proportion of posts containing YouTube videos over all posts averaged over time \\
$\overline{\rho_g}$ & Proportion of videos in group $g$ over all YouTube videos averaged over time \\

$T(t)$ & Indicator function, 0 when $t$ is before the treatment date and 1 otherwise \\
$d(t)$ & Function that returns the number of days since YouTube's announced treatment (i.e., 25 January), or 0 if $t$ is prior to the treatment date \\

\hline

\end{tabular}
\end{center}
\label{tab:variables}
\end{table}%

\section{Social Media Datasets and Time Series}

{While the section above describes our treatment and control groups for YouTube videos, here, we describe how we construct daily measures of sharing in each group for the ITS analysis (Table } \ref{tab:variables} {summarizes the notation of these measures).}
First though, we discuss the social media data from which we measure sharing of these videos across platforms.

\subsection{Social Media Datasets from Twitter and Reddit}

{As this paper examines cross-platform effects, we use} social media content collected from Twitter and Reddit to  extract links to YouTube videos and estimate daily sharing rates before and after this announcement.
For Twitter, we leverage an archive of tweets collected from Twitter's public sample stream, starting on 1 October 2018 and going until 31 May 2019 (representing four months on either side of YouTube's change). 
This dataset contains 827,334,770 tweets. 
Studies on this data source have identified shortcomings in its use for tracking topical coverage over time \cite{Morstatter2013}, but it should be sufficient for gauging changes in popularity of individual links, as suggested in the stream mining chapter of Leskovec et al. \cite{leskovec_rajaraman_ullman_2014}.
We similarly collect Reddit submissions using the PushShift.io collection \cite{baumgartner2020pushshift} during the same timeframe, resulting in 78,972,984 submissions.

We cannot directly search for links to YouTube within these datasets as link shorteners (e.g., bit.ly or TinyURL) obfuscate the terminal destination of a shortened link.
Such shorteners are often used in social media data, accounting for 5\% of links in our Twitter data and 0.03\% of links in our Reddit data. 
After applying the \texttt{urlexpander} package \cite{leon_yin_2018_1345144} to unshorten these links, we extract links to YouTube using the \texttt{youtube-data-api}  package \cite{leon_yin_2018_1414418}, as it captures both standard links to YouTube.com as well as shortened YouTube links (e.g., youtu.be) and embedded YouTube links.
The resulting dataset contains 3,658,455 and 4,363,124 links to YouTube on Twitter and Reddit, respectively. 
The \texttt{youtube-data-api} package also extracts YouTube video IDs from these links, yielding 3,273,473 and 3,641,872 unique YouTube videos on Twitter and Reddit.
Of these videos, 284,875 appear in both platforms, resulting in a total of 6,630,470 YouTube videos.

\begin{figure*}[hbt]
\begin{center}
\includegraphics[width=1\textwidth]{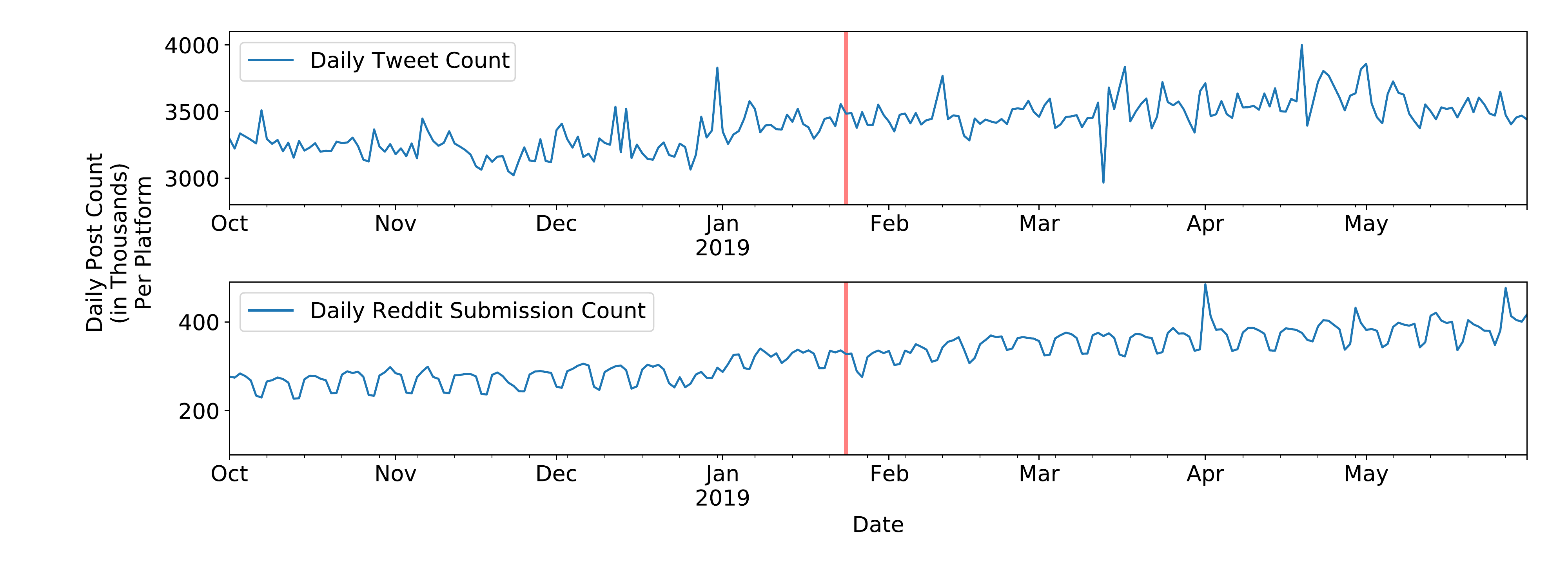}
\Description{Time series curves for both Twitter and Reddit, showing slightly increasing trends over the 8-month period, but no apparent impact of the January 25th announcement is discernible.}
\caption{\hl{Time series data on daily post rates in Twitter and Reddit. The red line marks January 25, the date of YouTube's announcement. Both platforms exhibit increasing trends in this data.}}
\label{fig:timeseries_post_count}
\end{center}
\end{figure*}

\begin{figure*}[bt]
\begin{center}
\includegraphics[width=1\textwidth]{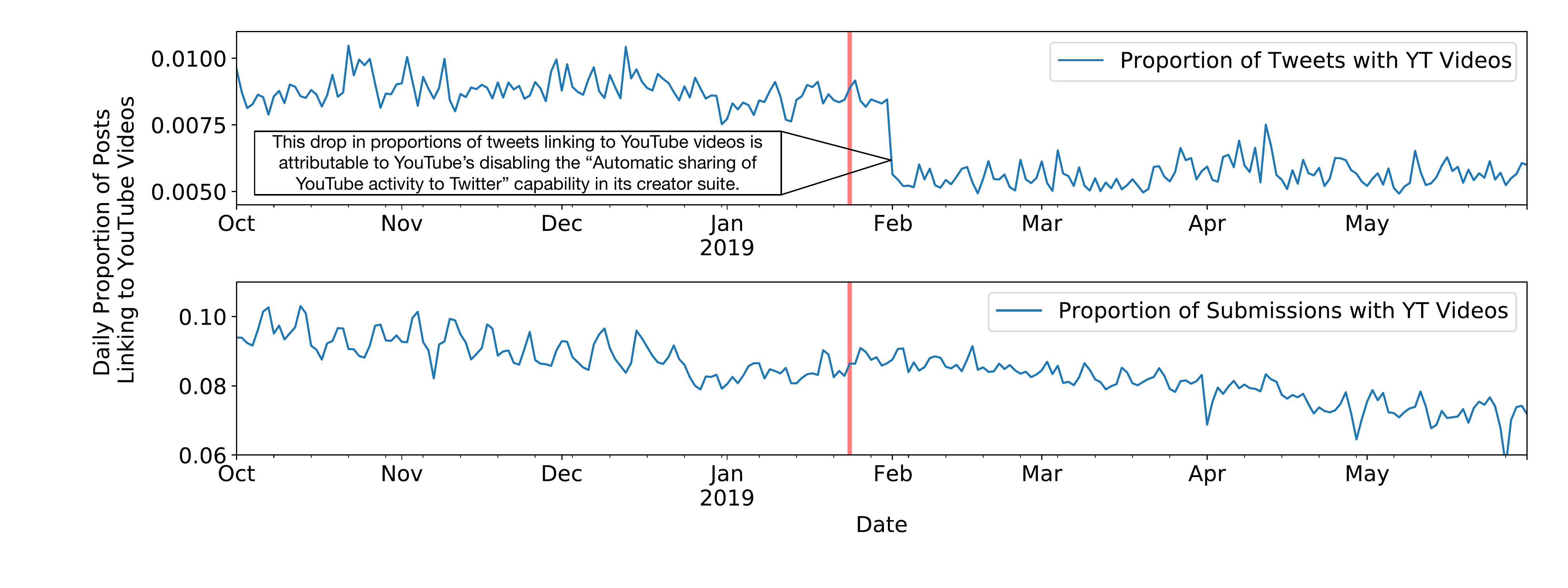}
\Description{Time series curves for proportions of posts in Twitter and Reddit, showing slightly decreasing trends over the 8-month period. In Twitter, we see a significant drop in proportion of posts linking to YouTube on February 1, an effect attributable to an engineering change at YouTube.}
\caption{\hl{Time series data on daily \emph{proportions} of posts in Twitter and Reddit. The red line marks January 25, the date of YouTube's announcement. A significant decline occurs in the proportion of tweets containing links to YouTube videos on 1 February, which is attributable to a change in YouTube's content creator suite that removed the ``Automatic sharing of YouTube activity to Twitter'' capability.}}
\label{fig:timeseries_prop_count}
\end{center}
\end{figure*}

{From this Twitter and Reddit data, we extract} daily counts of posts $n_t$ {as shown in Figure} \ref{fig:timeseries_post_count}.
From this figure, one can see the number of tweets on Twitter varies from three to four million per day with a slight upward trend.
For Reddit, submissions per day varies between 200 thousand and 500 thousand with a strong weekly periodicity and also exhibits an upward trend. 
{To account for these trends, we count those posts containing links to YouTube videos $v_t$ and divide by the number of posts per day in each platform; this measure yields a proportion of posts that contain links to YouTube $\rho_t = v_{t} / n_{t}$, shown in Figure} \ref{fig:timeseries_prop_count}.
Examining these proportions of posts containing links to YouTube, a different trend emerges: Both Twitter and Reddit see a decreasing trend in posts with links to YouTube over this 8-month timeframe.
{This downward trend is a confounder for analyses that only compare mean proportions before and after YouTube's announcement, as the mean proportions may appear lower after YouTube's announcement because of these trends.}
{Likewise, } a distinct drop in proportion of tweets containing links to YouTube is apparent on 1 February 2019.
This significant drop can be attributed to a change to YouTube's content publishing platform that went into effect on this day.
According to a support notice posted by YouTube's engineering team,\footnote{\url{https://support.google.com/youtube/thread/989408?hl=en}} YouTube had removed the ``Automatic sharing of YouTube activity to Twitter'' feature from content creators' ``Connected apps'' options, removing a pathway through which content creators could automatically share to Twitter when they uploaded a new video.
{These findings motivate the need for our ITS models, as these models allow us to disentangle level changes (e.g., YouTube's engineering change) and changes in trend that are attributable to a treatment at a particular moment in time.}

\subsection{Treatment and Control Time Series Data}

{We now focus on constructing time series data for our treatment and control groups $g$.}
{First, we divide the} time series for each video group into two regimes: pre- and post-YouTube's announcement, or 1 October 2018 to 24 January 2019 (inclusive) and 25 January 2019 to 31 May 2019.
{Then, we measure counts of links to each video group $v_{g,t}$, or} 1) links to conspiracy-labeled videos $v_{c,t}$, 2) links to videos from conspiracy-oriented channels $v_{cc,t}$, 3) links to AIN-produced videos $v_{a,t}$, and 4) links to videos from mainstream news sources $v_{m,t}$ on day $t$. 
{As with general YouTube sharing above, } we normalize these time series by calculating {group sharing proportions as the ratio of YouTube links to videos in this group} over the daily count of YouTube links on a platform: $\rho_{g,t} = v_{g,t} / v_{t} \forall g \in \{c, cc, a, m\}$. 

{For the pre- and post-treatment timeframes, comparing each group's mean proportion $\overline{\rho_g}$ in these two regimes provides insight into the coarse impact YouTube's intervention has had, as shown in Table} \ref{tab:yt_changes}.
{This table} demonstrates that changes in these proportions are small, often fractions of a percent. 
While the factors influencing these changes are varied, from YouTube's internal changes to content creator response to seasonality, we see consistent decreases in general YouTube sharing, conspiracy-labeled video sharing, and AIN sharing.
At the same time, we see increases in content from conspiracy channels and mainstream news sharing in both platforms.

\begin{table*}[h!]
\caption{Changes in YouTube sharing before and after YouTube's announcement, operationalized as the proportion of posts containing links to YouTube and the proportion of YouTube shares linking to conspiracy, AIN, and mainstream news videos. Mean changes in these proportions show decreases in sharing of general YouTube videos, conspiracy-labeled videos, and AIN-videos, whereas sharing of conspiracy channel-sourced videos and mainstream news videos are increasing.}
\small
\begin{center}
\begin{tabular}{l r r r c r r r}

\hline
 & \multicolumn{3}{c}{\textbf{Twitter}}  & &  \multicolumn{3}{c}{\textbf{Reddit}} \\ \cline{2-4} \cline{6-8}
Sharing Type & Pre- & Post-  & $\Delta$ & & Pre- & Post- & $\Delta$\\ \hline

General YouTube Sharing $\overline{\rho}$ & 0.8806\% & 0.5792\% & --34.23\% & & 8.969\% & 7.959\% & --11.25\% \\
Conspiracy Video Sharing $\overline{\rho_{c}}$ & 0.1824\% & 0.1438\% & --21.16\% & & 0.1481\% & 0.0872\% & --41.12\% \\
Conspiracy Channel Sharing $\overline{\rho_{cc}}$ & 0.2058\% & 0.2658\% & 29.16\% & & 0.1907\% & 0.2008\% & 5.296\% \\
AIN Channel Sharing $\overline{\rho_{a}}$ & 0.1336\% & 0.1265\% & --5.314\% & & 0.3407\% & 0.2761\% & -18.96\% \\
MSM Channel Sharing $\overline{\rho_{m}}$ & 0.2104\% & 0.3065\% & 45.67\% & & 0.4414\% & 0.5259\% & 19.14\% \\

\hline

\end{tabular}
\end{center}
\label{tab:yt_changes}
\end{table*}%

\begin{figure*}[p]
\begin{center}
\includegraphics[width=1\textwidth]{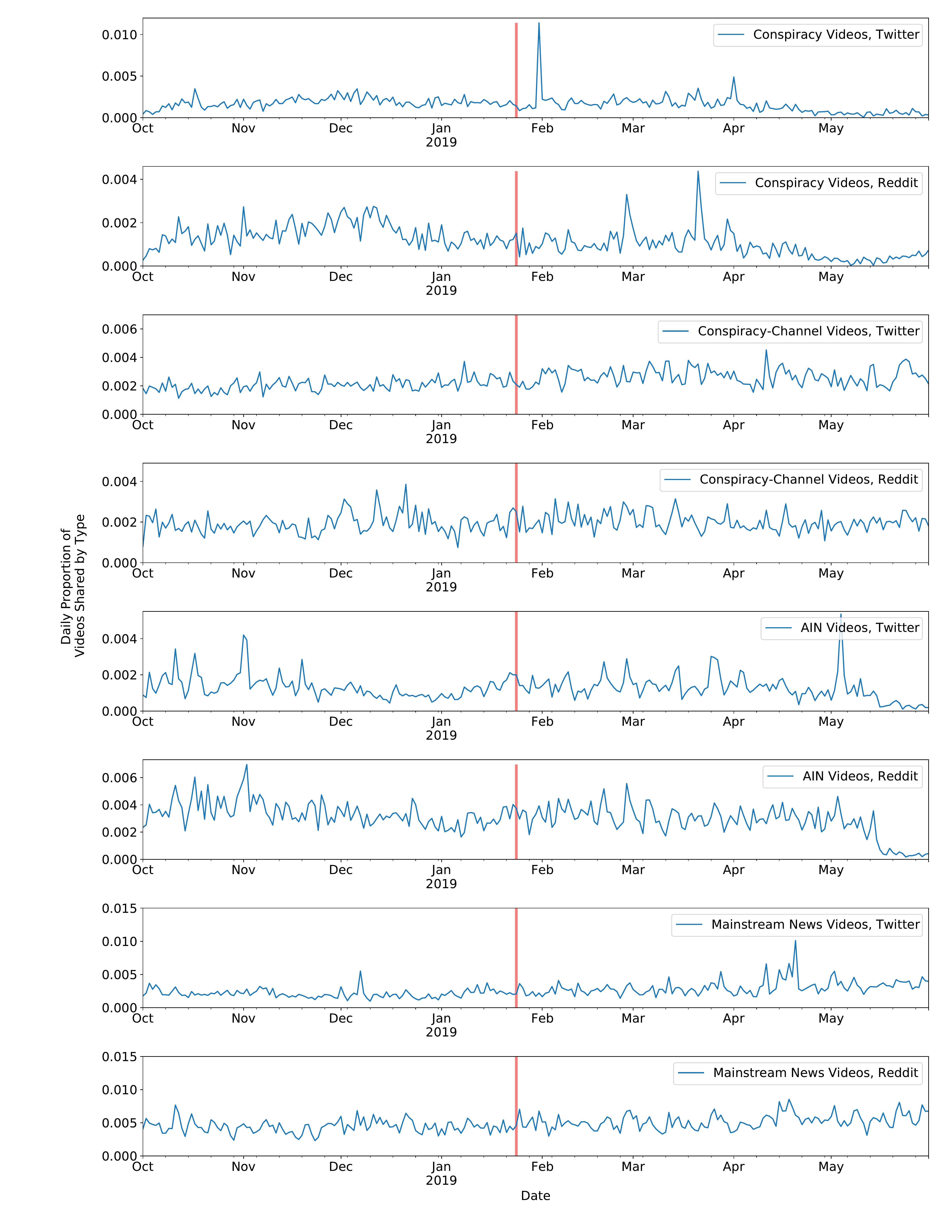}
\Description{Eight curves, representing the combinations of four video groups and two platforms. Conspriacy sharing seems to taper off in April, while AIN sharing appears to decrease in May. Conspiracy-channel and mainstream media content, however, seem to show little response.}
\caption{\hl{Time series data on daily proportions of videos shared that contain links to conspiracy-labeled, conspiracy-channel, AIN, and mainstream videos. The red line marks January 25, the date of YouTube's announcement. AIN sharing appears to taper off to a small amount in late May.}}
\label{fig:timeseries_types}
\end{center}
\end{figure*}

Relatedly, Figure \ref{fig:timeseries_types} shows the variations in proportions of sharing for the four {groups} we examine.
Qualitatively, this figure demonstrates that sharing of both Faddoul et al. conspiracy videos and AIN videos have decreasing tails towards the end of our timeframe. 
Conspiracy-video sharing shows a clear decrease in April, consistent with a decrease in in-platform recommendations found in Faddoul et al. \cite{Faddoul2020}.
AIN videos exhibit a similar decrease but not until the middle of May.
Conspiracy channel-sourced videos, however, do not exhibit a similar qualitative decline in either platform.
Mainstream news videos appear more prevalent on Reddit than Twitter and seem to be increasing in the proportion of sharing.
{The main implication from this analysis is that} the decline in conspiracy and AIN videos in the spring of 2020 suggests YouTube's actions may impact in sharing across platform boundaries, {but a more sophisticated analysis is needed}.

\section{Modeling YouTube's Impact in Pre- and Post-Treatment Sharing}

Given the difficulty in separating factors that contribute to changes in the aggregate sharing patterns shown in Table \ref{tab:yt_changes}, we introduce the comparative ITS log-log linear regression model to predict volumes of {our three treatment groups and one control group}.
In principal, this ITS model is similar to that employed by Chandrasekharan et al. \cite{Chandrasekharan:2017:YCS:3171581.3134666}, and more detail can be found in the ITS tutorial by Bernal et al. \cite{Bernal2017}.
ITS models are particularly applicable in this context as they allow us to capture both level and trend changes brought about by an intervention (here, YouTube's change).
Because platform effects and time lags likely impact the cross-platform sharing (e.g., a de-recommended YouTube video may be on the front page of Reddit the day before YouTube institutes its change, leading Reddit users to see this video for some period of time after YouTube's change), we expect the effects of YouTube's intervention to intensify over time, and the ITS model allows us to capture this effect with its trend parameter.
Furthermore, by adding a model fitted to control data, we can contextualize these results and increase our confidence that any effect observed only in the treatment set is a result of the intervention \cite{Bernal2018}.

{This ITS model is defined in a general form in Equation} \ref{eq:general_its_model}.
{At a high level, this model predicts the log-transformed number of videos in the target group on a given day as a function of the overall number of videos shared on the platforms, how often videos in this group were shared on the previous day, whether the day is before or after YouTube's announcement, and how many days have passed since this announcement.}
More specifically, $v_{t}$ captures the number of videos posted on day $t$, $v_{g,t-1}$ captures the lagged auto-regressive properties of sharing (i.e., the influence of the prior day in sharing) of this particular group $g$, $T(t)$ indicates whether the day $t$ is after 24 January 2019, $d(t)$ denotes the number of days after 25 January 2019 (and is equal to zero prior to that date), and $\beta_i$ variables are the coefficients we learn for each factor.

Sharing volumes are log-transformed to account for the highly skewed nature of sharing in social networks, leading us to expect proportional changes in response rather than directly linear changes.
Furthermore, we focus only on the \emph{percentage on videos} shared on each platform rather than the percentage of \emph{posts} because overall post volumes are subject to numerous additional factors {(e.g., YouTube's engineering change on 1 February)}.
Instead, our focus is on the \emph{distribution of videos} since YouTube is not decreasing the number of videos it recommends.

\begin{equation}
\label{eq:general_its_model}
\ln (v_{g,t} + 1) = \beta_1 \ln (v_{t} + 1) + \beta_2 \ln (v_{g,t-1} + 1) + \beta_3 T(t) + \beta_4 d(t)
\end{equation}

%
%
%
%
%
%

{Based on these models and the expectation that videos labeled as conspiracy and produced by AIN channels are subject to YouTube's de-recommendation treatment, then one should expect the following:}
First, assuming that YouTube's internal changes have a compounding impact over the broader ecosystem, we expect that $\beta_4 < 0$ in both Twitter and Reddit for our conspiracy-labeled and AIN-produced video groups.
{Conversely, for mainstream news, we expect no suppressive effect from YouTube's treatment, so $\beta_3 \ge 0$ and $\beta_4 \ge 0$ should hold.
Second, as YouTube's announcement suggests a phased deployment of their treatment, we anticipate no significant suppressive effect on the overall level of sharing of for each comparison group.
Lastly, for videos produced by conspiracy-oriented channels, we expect either a significant suppressive effect or no effect (i.e., $\beta_4 \le 0$).
}

\subsection{A Diagnostic ITS Model of YouTube Recommendations}

{
Prior to evaluating models results for our various video groups across platform boundaries, we first leverage daily recommendation data provided by Faddoul et al. to evaluate whether our ITS model can capture expected results for conspiracy videos \emph{specific to YouTube's platform}.
In this context, YouTube's announced intervention should manifest as a significant decline in the proportions of conspiracy videos YouTube recommends.
This test should validate our model construct since Faddoul et al. presents good evidence that conspiracy videos in YouTube have experienced this decline.
This experiment is constrained to conspiracy videos, however, as Faddoul et al.'s recommendation seeding method means we lack in-YouTube recommendation data seeded around our other video groups.
}

{
Our general ITS model in Eq.} \ref{eq:general_its_model} {is equally applicable to this context as the Faddoul et al. dataset includes lists of daily recommendations.
We can therefore define $v_t$ as the number of videos recommended on day $t$, and $v_{g,t}$ as the number of times videos in group $g$ have been recommended on day $t$.
Daily recommendation data in Faddoul et al. is incomplete, however, as several days appear missing from the data, so we apply a three-day moving average to smooth these missing data points.
}
{Table} \ref{tab:in_yt_model_results} {shows these YouTube-specific results, and as anticipated, the distance from treatment factor is significant and negative.}
{Hence, this modeling construct can detect this suppressive effect in YouTube, where we have high confidence that it exists.}

\begin{table*}[htp]
\caption{{Capturing impact of YouTube's announcement on conspiracy video recommendation within YouTube via a log-log linear regression ITS model. The model shows a decline in trend for these videos, consistent with Faddoul et al.}}
\small
\begin{center}
\begin{tabular}{l l r r}

\hline
& & \multicolumn{2}{c}{\textbf{In-YouTube Recommendations}} \\
Predictor & & $\beta$ & Std. Err. \\ \hline

Recommendation Volume & $v_{t}$    &    $0.0760^{***}$    &    $0.020$     \\

Lagged Conspiracy Recommendations & $v_{c,t-1}$    &    $0.8951^{***}$    &    $0.028$    \\

Treatment                        & $T(t)$    &    $0.0258$    &    $0.023$     \\

Distance from Treatment &   $d(t)$    &    $-0.0009^{**}$    &    $0.000$      \\
\hline
Observations &  & \multicolumn{2}{r}{237}    \\
$R^2$ &  & \multicolumn{2}{r}{1.000}    \\
\hline

\emph{Note:} & \multicolumn{3}{r}{$^*p<0.05; ^{**}p<0.01; ^{***}p<0.001$} \\ 
\hline

\end{tabular}
\end{center}
\label{tab:in_yt_model_results}
\end{table*}%

\subsection{Results for ITS Models of Pre- and Post-Treatment Sharing}

{After validating the ITS model within YouTube, we now turn to the cross-platform analysis and evaluate effects on sharing in Twitter and Reddit for our four comparison groups.}

\subsubsection{Conspiracy-Labeled Videos}

Table \ref{tab:con_model_results} shows results for our ITS model on conspiracy video sharing, showing the distance from YouTube's announcement $\beta_4$ is consistent with our expectation and manifests with a significant and negative effect on \emph{trends} for both Twitter and Reddit.
The level factor $\beta_3$ (being before or after the treatment date) is not significant in either platform, likewise consistent with our expectation of a phased roll-out of YouTube's intervention.
{This model suggests that, for each day that passes since YouTube's announcement, we can expect on average $0.7\%$ fewer conspiracy-labeled videos shared on both platforms.}

\begin{table*}[htp]
\caption{Capturing impact of YouTube's announcement on Faddoul et al. conspiracy video sharing in Twitter and Reddit with a log-log linear regression ITS model. Models show a consistent reduction in sharing trends significantly correlated to the distance from YouTube's announcement (i.e., the treatment) in both Twitter and Reddit}
\small
\begin{center}
\begin{tabular}{l l r r c r r}

\hline
& & \multicolumn{2}{c}{\textbf{Twitter}}  & &  \multicolumn{2}{c}{\textbf{Reddit}} \\ \cline{3-4} \cline{6-7}
Predictor & & $\beta$ & Std. Err.  &  & $\beta$ & Std. Err. \\ \hline

YouTube Volume & $v_{t}$    &    $0.2303^{***}$    &    $0.022$    & &    $0.1843^{***}$    &    $0.0200$    \\

Lagged conspiracy sharing & $v_{c,t-1}$    &    $0.3997^{***}$    &    $0.058$    & &    $0.4815^{***}$    &    $0.0560$    \\

Treatment                        & $T(t)$    &    $0.1246$    &    $0.078$    & &    $0.0586$    &    $0.085$    \\

Distance from Treatment &   $d(t)$    &    $-0.0079^{***}$    &    $0.001$    & &    $-0.0059^{***}$    &    $0.001$    \\
\hline
Observations &  & \multicolumn{2}{r}{242}  & &  \multicolumn{2}{r}{242} \\
$R^2$ &  & \multicolumn{2}{r}{0.971}  & &  \multicolumn{2}{r}{0.985} \\
\hline

\emph{Note:} & \multicolumn{6}{r}{$^*p<0.05; ^{**}p<0.01; ^{***}p<0.001$} \\ 
\hline

\end{tabular}
\end{center}
\label{tab:con_model_results}
\end{table*}%

\subsubsection{AIN Videos}

Table \ref{tab:ain_model_results} presents results for AIN video sharing in Twitter and Reddit.
As with conspiracy videos and our expectation that YouTube's changes will decrease sharing beyond its boundaries, distance from YouTube's announcement $\beta_4$ again manifests with a significant and negative effect on \emph{trends} in AIN sharing for both Twitter and Reddit.
{This model suggests that, for each day that passes since the announcement, we can expect approximately $0.3\%$ fewer AIN videos shared on both platforms.}

\begin{table*}[htp]
\caption{Capturing impact of YouTube's announcement on the sharing of AIN-produced videos in Twitter and Reddit with a log-log linear regression ITS model. Model results show a consistent reduction in AIN sharing trends significantly correlated to the distance from YouTube's announcement (i.e., the treatment) in both Twitter and Reddit.}
\small
\begin{center}
\begin{tabular}{l l r r c r r}

\hline
& & \multicolumn{2}{c}{\textbf{Twitter}}  & &  \multicolumn{2}{c}{\textbf{Reddit}} \\ \cline{3-4} \cline{6-7}
Predictor & & $\beta$ & Std. Err.  &  & $\beta$ & Std. Err. \\ \hline

YouTube Volume & $v_{p,t}$    &    $0.1330^{***}$    &    $0.018$    & &    $0.1073^{***}$    &    $0.019$    \\

Lagged AIN sharing & $a_{p,t-t}$    &    $0.6209^{***}$    &    $0.051$    & &    $0.7554^{***}$    &    $0.042$    \\

Treatment                        & $T_t$    &    $0.0784$    &    $0.076$    & &    $0.1237$    &    $0.063$    \\

Distance from Treatment &   $d_t$    &    $-0.0036^{***}$    &    $0.001$    & &    $-0.0033^{***}$    &    $0.001$    \\
\hline
Observations &  & \multicolumn{2}{r}{242}  & &  \multicolumn{2}{r}{242} \\
$R^2$ &  & \multicolumn{2}{r}{0.988}  & &  \multicolumn{2}{r}{0.995} \\
\hline

\emph{Note:} & \multicolumn{6}{r}{$^*p<0.05; ^{**}p<0.01; ^{***}p<0.001$} \\ 
\hline

\end{tabular}
\end{center}
\label{tab:ain_model_results}
\end{table*}%

\subsubsection{Conspiracy-Channel Videos}

The model of YouTube's announcement on the sharing of videos from conspiracy-oriented channels follows a different pattern that what we find in conspiracy-labeled and AIN-produced videos.
Table \ref{tab:con_chan_model_results} suggests that, unlike above, YouTube's announcement has had no significant impact on \emph{trends} in sharing of conspiracy videos on either platform.
Surprisingly, the overall \emph{level} of conspiracy sharing increases significantly on Reddit during the post-treatment regime.
The model suggests that, after YouTube's announcement, sharing of videos from conspiracy-oriented channels has increased by approximately $16\%$ on Reddit.

\begin{table*}[htp]
\caption{Capturing impact of YouTube's announcement on sharing of videos from conspiracy-oriented channels in Twitter and Reddit with a log-log linear regression ITS model. Models show an inconsistency between Twitter and Reddit wherein conspiracy sharing on Reddit sees an increase following the treatment of YouTube's announcement, but this effect is not significant in Twitter.}
\small
\begin{center}
\begin{tabular}{l l r r c r r}

\hline
& & \multicolumn{2}{c}{\textbf{Twitter}}  & &  \multicolumn{2}{c}{\textbf{Reddit}} \\ \cline{3-4} \cline{6-7}
Predictor & & $\beta$ & Std. Err.  &  & $\beta$ & Std. Err. \\ \hline

YouTube Volume & $v_{t}$    &    $0.2977^{***}$    &    $0.025$    & &    $0.3169^{***}$    &    $0.024$    \\

Lagged cons. chan. sharing & $v_{cc,t-1}$    &    $0.2506^{***}$    &    $0.062$    & &    $0.1696^{**}$    &    $0.062$    \\

Treatment                        & $T(t)$    &    $0.0472$    &    $0.040$    & &    $0.1627^{**}$    &    $0.049$    \\

Distance from Treatment &   $d(t)$    &    $-0.0003$    &    $0.000$    & &    $-0.0008$    &    $0.001$    \\
\hline
Observations &  & \multicolumn{2}{r}{242}  & &  \multicolumn{2}{r}{242} \\
$R^2$ &  & \multicolumn{2}{r}{0.998}  & &  \multicolumn{2}{r}{0.997} \\
\hline

\emph{Note:} & \multicolumn{6}{r}{$^*p<0.05; ^{**}p<0.01; ^{***}p<0.001$} \\ 
\hline

\end{tabular}
\end{center}
\label{tab:con_chan_model_results}
\end{table*}%

\subsubsection{Mainstream News Videos}

{Results of our ITS models above show declines in Twitter and Reddit sharing of particular anti-social video types following YouTube's announcement.}
{To ensure these results are specific to anti-social video types}, this section turns {videos produced by mainstream media outlets, as these videos should not fall into YouTube's description of ``harmful or potentially misinforming'' content}.
Table \ref{tab:msm_model_results} shows no significant decrease in either the level or trend in sharing of mainstream news videos on either Twitter or Reddit between the pre- and post-treatment regimes.
{On the contrary, mainstream news has seen significant \emph{increases} in both platforms; Twitter sees progressively more mainstream YouTube video sharing, and Reddit sees an increase in both the overall level and trend in mainstream sharing.}

\begin{table*}[htp]
\caption{Capturing impact of YouTube's announcement on mainstream news video sharing in Twitter and Reddit with a log-log linear regression ITS model. The models show sharing of mainstream news videos in both platforms actually has increased in both platforms.}
\small
\begin{center}
\begin{tabular}{l l r r c r r}

\hline
& & \multicolumn{2}{c}{\textbf{Twitter}}  & &  \multicolumn{2}{c}{\textbf{Reddit}} \\ \cline{3-4} \cline{6-7}
Predictor & & $\beta$ & Std. Err.  &  & $\beta$ & Std. Err. \\ \hline

YouTube Volume & $v_{t}$    &    $0.2611^{***}$    &    $0.024$    & &    $0.3214^{***}$    &    $0.027$    \\

Lagged MSM sharing & $v_{m,t-1}$    &    $0.3438^{***}$    &    $0.061$    & &    $0.3068^{***}$    &    $0.059$    \\

Treatment                        & $T(t)$    &    $-0.0506$    &    $0.053$    & &    $0.0971^{*}$    &    $0.043$    \\

Distance from Treatment &   $d(t)$    &    $0.0023^{**}$    &    $0.001$    & &    $0.0012^{*}$    &    $0.001$    \\
\hline
Observations &  & \multicolumn{2}{r}{242}  & &  \multicolumn{2}{r}{242} \\
$R^2$ &  & \multicolumn{2}{r}{0.996}  & &  \multicolumn{2}{r}{0.998} \\
\hline

\emph{Note:} & \multicolumn{6}{r}{$^*p<0.05; ^{**}p<0.01; ^{***}p<0.001$} \\ 
\hline

\end{tabular}
\end{center}
\label{tab:msm_model_results}
\end{table*}%

\subsection{Results Summary}

Table \ref{tab:model_results_summary} summarizes results from the above models, showing that both conspiracy-labeled and AIN-authored videos experience significant downward trends in sharing on both platforms.
Furthermore, as the mainstream media control set experiences significant increases in trend across both platforms and a significant level change in Reddit, the control suggests trend changes experienced by conspiracy-labeled and AIN videos is differentiated from overall changes in news sharing on YouTube or its popularity.

\begin{table*}[htb]
\caption{Summary of Trend and Level Changes Across Video Types and Platforms. $\varnothing$ represents no significant effect, and we use $+/-$ to denote significant increases or decreases in this factor.}
\small
\begin{center}
\begin{tabular}{l l r r c r r}

\hline
& & \multicolumn{2}{c}{\textbf{Twitter}}  & &  \multicolumn{2}{c}{\textbf{Reddit}} \\ \cline{3-4} \cline{6-7}
Video Type & & Trend & Level  &  & Trend & Level \\ \hline

Conspiracy-Labeled Videos &     &    $-^{***}$    &    $\varnothing$    & &    $-^{***}$    &    $\varnothing$    \\

AIN-Channel Videos             &     &    $-^{***}$    &    $\varnothing$    & &    $-^{***}$    &    $\varnothing$    \\

Conspiracy-Channel Videos &     &    $\varnothing$    &    $\varnothing$    & &    $\varnothing$    &    $+^{**}$    \\

MSM-Channel Videos           &     &    $+^{**}$    &    $\varnothing$    & &    $+^{*}$    &    $+^{*}$    \\
\hline

\emph{Note:} & \multicolumn{6}{r}{$^*p<0.05; ^{**}p<0.01; ^{***}p<0.001$} \\ 
\hline

\end{tabular}
\end{center}
\label{tab:model_results_summary}
\end{table*}%

\section{Discussion}

Apparent from these studies {is that} YouTube's announcement and resulting change to their recommendation algorithm {coincide with significant declines in} the sharing of videos labeled as conspiracy content (see Table \ref{tab:con_model_results}) and videos produced by alternative-information channels (Table \ref{tab:ain_model_results}).
{By comparing these declines with} the relative increase in mainstream news sharing during this same timeframe, we have evidence {supporting the claim that YouTube's actions have suppressive effects on particular kinds of content beyond its boundaries}.
{While potential causal claims must be tempered by the many confounders in this space,} we discuss a selection of implications stemming from these results.

\subsection{De-Recommendation Versus Deletion and Anti-Censorship}

A key point in ``de-recommendation'' as a tool for platform moderation and governance is how it is differentiated from the deletion of content.
While removing content from YouTube's platform may be the most heavy-handed way to reduce exposure to this harmful and misinforming content, it is clearly not the only solution, so the question becomes one of which action best achieves the desired effect of increasing information quality.
On the one hand, as shown in Chandrasekharan et al., wholesale banning and deletion of anti-social users and communities is effective within a specific platform \cite{Chandrasekharan:2017:YCS:3171581.3134666}, and if YouTube were to delete such content, as they have announced with white-supremacist videos, {an individual's unintentional} stumbling across links to this content from other platforms (e.g., Twitter or Reddit) are of little value or risk.
On the other hand, actively deleting content, especially if that content is popular, has the potential to backfire and counter-productively increase the target content's popularity, a phenomenon dubbed the ``Streisand effect'' \cite{Jansen2015}.

Platform-level deletion of popular but subversive content may be particularly likely to trigger the Streisand effect given current anti-censorship concerns professed by political and media elites in the US.
AIN channels in particular may push such anti-censorship narratives when platforms delete their content, as a consistent theme among these channels is their anti-establishment and anti-mainstream position.
Thus, when a platform like YouTube actively deletes content, these channels may respond with claims of censorship and galvanize their followings.

\begin{figure*}[h!]
\begin{center}
\includegraphics[width=0.75\textwidth]{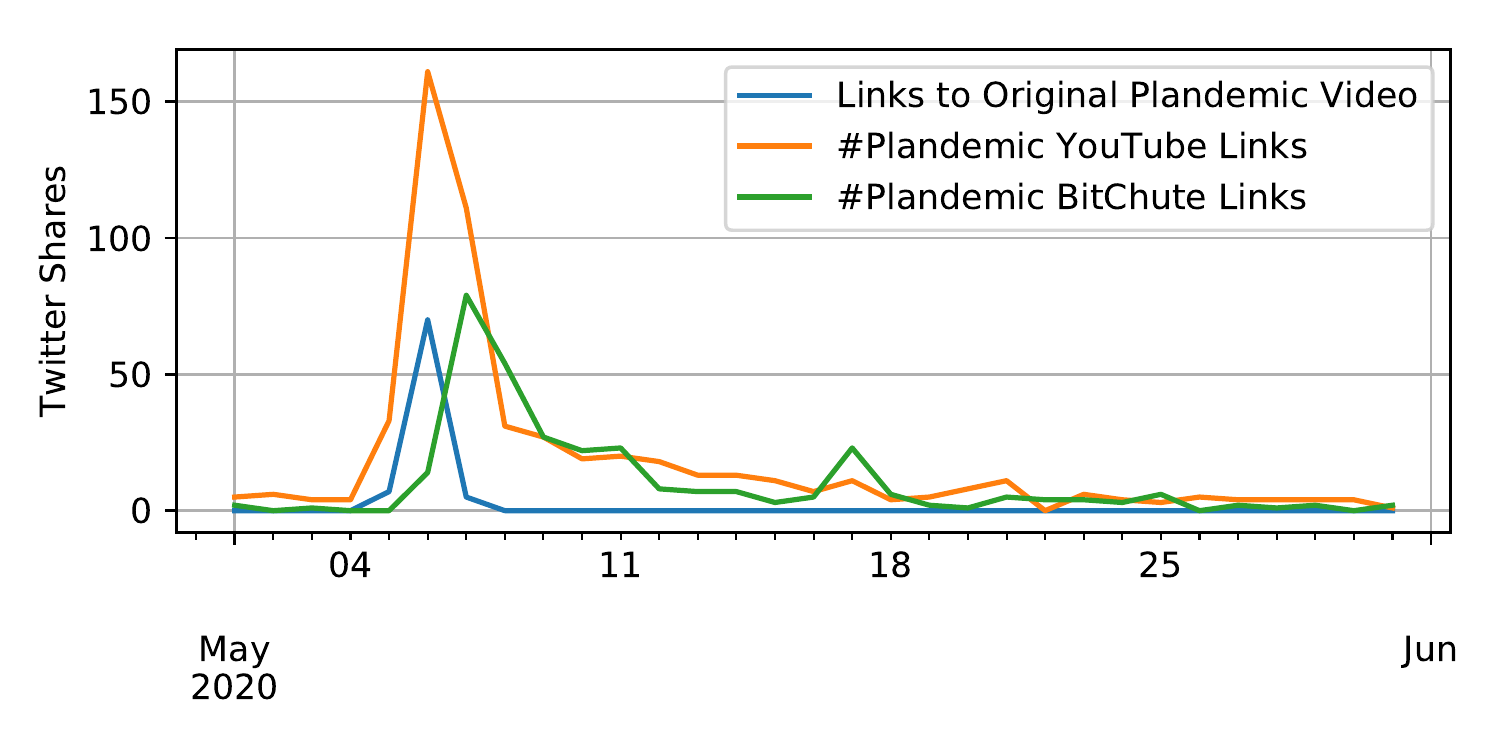}
\Description{Time series data on daily posting of tweets that link to the original Plandemic YouTube video, tweets mentioning Plandemic and linking to YouTube, and tweets mentioning Plandemic and linking to BitChute. We see a spike in links to the original Plandemic YouTube video that is dwarfed by all links to YouTube mentioning Plandemic. We also see a lagged effect in links to BitChute, which occur after YouTube removed the original video.}
\caption{{Time series data on daily posting of tweets that link to the original Plandemic YouTube video (blue curve); tweets mentioning Plandemic and linking to YouTube (orange curve); and tweets mentioning Plandemic and linking to BitChute, an alternative video-sharing site (green curve). YouTube removed the original Plandemic video on May 5, with a related drop in links to that specific video ID on the 6th. Following the initial removal, individuals continued to re-post the video to YouTube and transitioned to sharing the video on BitChute.}}
\label{fig:timeseries_plandemic}
\end{center}
\end{figure*}

A prime and recent example of backfires in response to platform deletion is with the popularity of the ``Plandemic'' movie, a debunked and scientifically inaccurate anti-vaccine video that has been extremely popular during the recent COVID-19 pandemic \cite{Neil2020}.
As Lytvynenko discusses in the popular press article about the rise of this video \cite{Wakabayashi2019}, the main personality in ``Plandemic'' is framed as a whistleblower, and the resulting deletion of this video on Facebook and YouTube bolster the anti-censorship framing.
{This response is clearly visible in Figure} \ref{fig:timeseries_plandemic}, {which shows the tweets per day that link to the original Plandemic YouTube video, links to other Plandemic-related YouTube content (much of which were reposts of the original video following its removal), and links to the Plandemic video on BitChute, an alternative, ``moderation-lite'' video-sharing platform} \cite{10.1145/3372923.3404833}.
{Figure} \ref{fig:timeseries_plandemic} {shows a burst in sharing of the original Plandemic video on 4 May 2020, peaking on 6 May, before dropping back to zero following YouTube's removal of the original video for violations.}
{Despite YouTube's deleting this video, the figure shows longer-lived sharing of reposted Plandemic videos for the rest of the month of May.}
{Had YouTube instead de-recommended this video, the multitude of video reposts and transition to the more hate-filled BitChute } \cite{10.1145/3372923.3404833} {may not have occurred.}

Consequently, de-recommendation could be an alternate response that social media platforms could take without actively deleting this video and thereby avoid triggering this anti-censorship framing.
Prior research on the dissemination of ``fake news'' \cite{Grinberg2019} suggests low-quality content tends to stay within insular communities, so removing pathways that incidentally expose the uninitiated (e.g., via automated recommendations) may therefore still achieve the pro-social outcomes we see above.

\subsection{De-Recommendation Versus Deletion and Transparency}

While deletion may be the most effective strategy to mitigate YouTube's role in spreading misinforming and harmful content online, making YouTube responsible for policing what is considered ``harmful'' discourse introduces its own difficulties.
Since YouTube does not provide a mechanism through which the public can identify this harmful content, an individual is given little feedback if she inadvertently stumbles upon such a video (e.g., from a link on Facebook, WhatsApp, or Twitter).
Likewise, a content creator whose videos are flagged by this system has no way of knowing and therefore no recourse in challenging YouTube's classification. 

{This limited transparency in what content is subject to YouTube's de-recommendation sanctions is particularly concerning given the volume of political speech on the platform.}
{In the current context, the public must trust that YouTube is not exercising this sanction against content that it might consider harmful to its business model (e.g., videos promoting regulation on social media platforms or online advertising restrictions) but is otherwise legitimate.}
{Twitter and Facebook, in contrast, label violative content or posts found to be dishonest or misleading and are thus significantly more transparent.}
{While an operational security argument could be made regarding YouTube's hesitance in being forthcoming about how these sanctions are applied, such arguments must be balanced against the public good and YouTube's dominant role in the online information space.}
Given YouTube's position as the dominant video sharing platform, such lack of transparency, and the absence of regulatory incentives to support the public good, making YouTube a major arbiter of what is considered allowable concentrates significant political and informational power into a single entity.

\subsection{Growing Content from Conspiracy-Oriented Channels}

An additional difficultly particularly germane to this discussion is the ambiguity in defining ``misinforming and harmful'' content.
In our study, YouTube's action does not appear to have significant impact on content from conspiracy-oriented channels; in fact, we see, as with the control set, an \emph{increase} in conspiracy-channel sharing. 
While it could be that the propagation of conspiracy-channel videos is entirely separate from YouTube's recommendations, we instead propose two other explanations:

One plausible explanation may lie in the content creation strategies employed by these channels. 
YouTube's announcement, regardless of underlying platform changes, may incentivize outlets to shift their content creation strategies to be more inline with YouTube's policies.
To that end, channels may strategically remove content that could be considered harmful or misinforming under the quality guidelines Google has made available.
In fact, we see some evidence for this activity in Table \ref{tab:video_overlap}, where many conspiracy-labeled videos were not acquired in our channel-directed collection, as they had been deleted or made private.

Another more troubling explanation for the growth in content from conspiracy-oriented channels stems from what \emph{replaces} the content YouTube is no longer recommending.
Holding other quantities equal, removing harmful videos from recommendation opens slots for new videos to be recommended.
It is therefore possible that the audience to which these harmful videos would be recommended are instead recommended the more innocuous or borderline videos from these channels, as these conspiracy videos may be semantically or socially adjacent to the harmful or misinforming content.
That is, this audience may have then inadvertently been diverted from ``harmful'' content to other content from these conspiracy channels.
Without additional insight from YouTube about how this mechanism works and whether harmful videos are replaced with the next most similar video or with a more pro-social video, we are limited in understanding this interplay.
It is nonetheless important to consider these unintended consequences and whether YouTube's actions create new incentives for more subtle misinformation and harmful content.

\section{Threats to Validity and Other Limitations}

We acknowledge the difficulty in making causal claims in this context, as the information ecosystem is complex, and we are able to capture only portions of it.
{For instance, our work integrates only online behaviors in a few platforms, but the modern information space covers online sources, traditional cable and written sources, and real-world interactions.}
{Consequently, effects observed above may be caused by external factors beyond what we measure.}
Below, we identify some potential threats to {the validity of} our claims, {alternate explanations, and} related limitations.

\subsection{Views Versus Shares}

{Likely the largest concern with the analysis presented above is its focus on information production rather than \emph{consumption}, as our analysis omits analysis of video viewership.}
{As a consequence, while our results may be internally consistent in showing specific types of videos are \emph{shared} less, our analysis does not preclude the possibility that such videos receive as many views or more than they did prior to YouTube's intervention.}
The best solution to this {external} threat would be to {replace our use of shares with} impression and view data, {but} this information is highly protected by the platforms and is therefore difficult to collect.
{Collecting this information would generally require either collaboration with channel owners or with the platforms directly.}
{While YouTube makes engagement data available to content creators, some creators may be disincentivized to share such analytics given the nature of this research.}
{Likewise, platforms guard large-scale engagement data to protect advertising interests, and such collaborations bring their own concerns about impartiality and objective research}.
{Facebook has shown an openness to allowing researchers access to this data without publication restrictions via Social Science One,\footnote{\url{http://socialscience.one/}} its joint effort with Social Science Research Council, but neither Twitter nor Reddit nor YouTube have joined such efforts}.
{Alternatively, researchers could monitor daily view counts from a video's YouTube page, similar to the daily recommendation collection in Faddoul et al., but such approaches are intrinsically irreproducible since the data cannot be recollected.}
{Given these constraints}, we use link sharing as a proxy for consumption, {which is supported by previous empirical research on Twitter} \cite{Chaintreau2016}.

We {also} note that we make an important distinction between the \emph{proportion of shared links} and the \emph{proportion of shared videos}.
Recommendation changes should affect this composition of shared videos under the expectation that changes to the composition of recommendations, as YouTube claims, will be reflected in changes to composition of what is shared, regardless of audience size.

\subsection{Identifying Potentially Harmful Videos for Treatment}

{
Related to our inability to measure views, another critical confounder in this paper concerns our selection process for creating treatment and control groups.
Since YouTube does not share this information, external researchers cannot know precisely what videos comprise the treatment set (i.e., those videos to which de-recommendation has been applied), and the treatment groups we identify above are likely both incomplete and noisy. 
That is, we likely have untreated videos in our treatment groups (i.e., false-positives) and have missed videos that are subject to this treatment (i.e., false-negatives). 
While these issues are important, they are likely insurmountable without YouTube's insight, which they are unwilling to provide. }

{
For this study, however, these issue does not invalidate our results. 
Rather, type-I and type-II errors in the construction of our treatment groups will primarily dilute treatment effects, as these confounders affect our selection process. 
Consequently, coefficients identified in our models are likely \emph{stronger} than we measure. 
Specifically, as more untreated videos are added to our treatment groups (representing false-positives), changes in recommendation to the overall set become less prevalent, meaning we are less likely to see differences in cross-platform sharing brought about by YouTube's intervention. 
Likewise, as more treated videos are held out of our treatment groups, these false-negatives must end up either in our control groups or are not tested; in the former case, we will see less differences between treatment and control groups, and in the latter, we have smaller treatment groups, meaning effects must be larger to be significant. 
In either case, if effects remain significant despite these confounders, we can be confident these effects are valid findings (though the ultimate cause may still be elusive).
}

{
To validate this claim, we perform a post-hoc test on a subset of the Faddoul et al. dataset, removing the 28 videos posted by mainstream media organizations, as these videos are likely false-positives. 
Our expectation here is that the coefficient of the distance-to-treatment factor should be larger in magnitude than what we originally find.
In the original model shown above, we find $\beta_4 = -0.000916$; in this post-hoc model, we find $\beta_4 = -0.000937$.
Likewise, we find smaller a $p$ value in this post-hoc model ($0.007238 < 0.008232$).
Both of these results are consistent with the expectation that errors in classification will \emph{reduce} observed treatment effects, suggesting our results are more likely to underestimate the suppressive effects of YouTube's intervention.
As mentioned, transparency from YouTube would obviate these issues.
}

\subsection{Mainstream Media as a Control}

{Similar to concerns about classification in our conspiracy video group,} our selection of mainstream media channels as a control set may also threaten our results.
Guidance from in Bernal et al. is to select highly similar control sets while avoiding sets that may be subject to indirect effects from the intervention \cite{Bernal2018}, and we initially anticipated mainstream videos would not be subject secondary effects from de-recommending misinforming videos.
In retrospect, the increase we see in YouTube's mainstream news sharing on both Twitter and Reddit may result from indirect effects wherein YouTube replaces recommendations to misinforming content with recommendations to more mainstream content.
While our findings remain consistent in the decreasing trends in conspiracy-labeled and AIN videos even given this potential indirect effect on the control set, this decrease may stem from an overall decline in YouTube popularity that the indirect effects on the control mask.
Our finding of null effects on trends in conspiracy-channel videos provide some evidence that this alternate interpretation is incorrect.
An alternative control set could be devised, but it would likely need to deviate further from the type of videos under treatment.

\subsection{Unknown Lag in YouTube's De-Recommendation Interventions}

Our choice of January 25 to divide treatment regimes is also potentially problematic, as YouTube may not have deployed these changes at the same time as their announcement.
Results in Faddoul et al., for instance, show recommendations to conspiracy videos did not reach a minimum until late April 2019, with marked declines not obvious from their data until March \cite{Faddoul2020}.
Despite this concern, no other date presents a clearer distinction, as YouTube's blog post says their deployment will be a ``gradual change'' and will be rolled out over time as their models ``become more accurate'' \cite{YouTubeTeam2019}.
In fact, results shown in Suzor \cite{Suzor2019} show a marked decrease in AIN recommendations on 14 February, consistent with a gradual deployment and further suggesting that YouTube did not deploy this change on a single date.
ITS models account for these gradual deployments through the trend factor, however, and such methods are generally used in these instances where the treatment effect is expected to compound over time.
As such, while we acknowledge this potential issue in date selection, better results are unlikely without more input from YouTube about the exact date they deployed this change {or when a particular video became subject to this sanction}.

\subsection{Potential Alternative Explanations}

{As mentioned, other factors beyond what we measure may drive the effects we see beyond a clear causal relationship stemming from YouTube's de-recommendation actions.}
{Below, we outline a few potential alternative explanations, including other platform-level curation actions and changes in audience interests.}

{First, as noted above and shown in Figure} \ref{fig:timeseries_types}, declines in conspiracy and AIN sharing do not  occur simultaneously; rather, conspiracy declines in April, while AIN content declines in May.
{One possible explanation for this observation is that YouTube's de-recommendation sanctions were applied at different times to these groups.}
{Alternatively, our results} may stem from an additional YouTube announcement in early June that white supremacist content or ``videos that promote or glorify Nazi ideology'' are to be removed from the platform \cite{YouTubeTeam}. 
Though this more recent announcement states these new content policies went into effect on June 5, they mention ``it will take time for [YouTube's] systems to fully ramp up and [YouTube will] be gradually expanding coverage over the next several months'' \cite{YouTubeTeam}.
It is therefore possible that the decline in AIN-related content stems from removal as a precursor to the June 5 announcement. 
To address this potentiality, we have checked our initial set of 40,764 AIN-produced videos to see how many remain on YouTube's platform, and we find 37,612 of these 40,764 remained in July 2019; that is, only $7.732\%$ of videos in our AIN set have been removed.

{Related to YouTube's platform curation actions, while we attribute changes in conspiracy- and AIN-produced-video sharing in Twitter and Reddit to YouTube's actions, these results may be attributable to actions taken directly by these other platforms.}
{As Twitter and Reddit are known to take similar suppressive actions (e.g., labeling tweets or quarantining subreddits), it is possible that our selected videos are subjected to similar suppressive action within those platforms separately.}
{This possibility would require collaboration across the platforms, however, which seems unlikely given known disagreements between Facebook and Twitter about how they apply their community standards to content on their platforms.}
{At the same time, AIN channels in particular are well-known as sources of misinformation, so this possibility is worth considering.}
{While eliminating this possibility from our analysis is likely impractical as we would need sensitive input from all three platforms, future research relying on Facebook's Social Science One initiative could potentially provide some insight, as the data contained therein includes third-party fact-checking outcomes at the individual link level (which would include links to YouTube videos).}

{Platform actions aside, our results may capture changes brought about by YouTube's actions but miss the mechanism causing these changes, as the above results could be explained by large-scale changes in audience interests.}
{That is, if the audience of Internet users collectively shifted attention away from explicit conspiracy content or videos produced by alternative information sources, we might observe similar declines in sharing.}
{In that case, creators of conspiracy content may have reacted by producing other types of content that are more popular, thereby explaining why we see an increasing in conspiracy-channel sharing.}
{At the same time, content creators may perceive YouTube's actions as a signal of such a shift, and their desire to maintain income on the platform may strategic choices to highlight other content outside of these groups}.
{Munger and Phillips provide some evidence of this change in audience interests, as their analysis suggests interest in ``far-right'', alternative YouTube content peaked in 2017} \cite{Munger2019}.
{At the same time, recent studies by Pew finds popular distrust of mainstream news sources and fracturing of the news ecosystem is \emph{widening} rather than shrinking} \cite{Jurkowitz2020}, {which would suggest our findings of increased sharing among mainstream news videos is not driven by audience interests.}
{Closer collaborations with YouTube about which specific videos are subject to de-recommendation and \emph{when} that action occurred would better address these audience-level concerns.}

\subsection{Platform Selection, Collection, and Related Biases}

{While Twitter and Reddit are popular platforms in the US, our use of them to measure daily shares for videos in our treatment and control groups introduce key biases in their own rights.}
In particular, Facebook's exclusion from this analysis is a key limitation, as Facebook's massive user base and population of topic-specific pages and groups could be a gathering place for people interested in the potentially anti-social topics YouTube wishes to target.
{Facebook also has a large ecosystem of private groups and pages, where anti-social content may be shared, and as platforms take a firmer stances on such content, users may be incentivized to move such discussions into more private spaces.}
This issue is difficult to address given current data limitations from Facebook, but we are actively working on future research to address this limitation.

{Relatedly, our use of Twitter's public sample stream, which is only 1\% of the full conversation, may introduce its own biases based on what YouTube videos became popular on Twitter.}
{Relevant studies on Twitter's public sample stream, for example, have found it to be biased against less-popular content} \cite{Morstatter2013}.
{It is therefore possible that our results on Twitter are only applicable to popular YouTube videos; the replication of this result on Reddit reduce this concern to a degree but does not eliminate it.}
{One way to mitigate this concern would be to purchase data from Twitter during this timeframe, but purchasing \emph{all} content posted in this 8-month period would be prohibitively expensive; one could restrict this purchase to tweets including YouTube links, but this query would be confounded by hyperlink shortenings, as we discuss above.}

{Finally, these results have potential geographic restrictions.}
{Reddit's user base, in particular, is primarily US-centric, and the US comprises a significant proportion of Twitter's users, so our results may hold only in the US.}
{This US focus is compounded by the use of AIN channels, which are also generally specific to the US political sphere, leaving open questions about whether YouTube's de-recommendation affects sharing in other national contexts, where political and public-health misinformation may be especially prevalent.}
{The public health context is also excluded in our analysis, as we focus on conspiracy- and AIN-produced videos, which is especially critical in times of global pandemic.}
{This shortcoming stems primarily from limited datasets on YouTube-centric public health misinformation and whether these videos have been subjected to YouTube's de-recommendation, which is an area of future work.}

\subsection{Limitations on Reproducibility}

{Beyond these threats is a key limitation stemming from} our reliance on Faddoul et al. for a collection of conspiracy-focused videos, which {complicates reproducing similar analyses for other timeframes}. 
While Faddoul et al. do provide strong evidence for a set of videos that have been subjected to YouTube's de-recommendation, one cannot retrospectively reproduce this dataset as their methodology relies on streaming collections of YouTube recommendations.
{Again, the only path forward to conclusively resolve these concerns is likely to be partnering with YouTube and obtain a listing of videos YouTube has identified as conspiracy, but this approach would require platform-level buy-in and comes with its own set of concerns as noted previously.}

\section{Conclusions}

{Ultimately, this work shows YouTube's efforts to suppress potentially harmful and misinforming content on its own platforms coincide with reductions in sharing on Twitter and Reddit. }
{While causal claims are difficult to make in this complex environment and given the unknowns in what videos actually are subject to this de-recommendation treatment, these results support the hypothesis that in-platform suppression, without deletion, has suppressive effects across other online spaces.}
While it is unlikely similar de-recommendation approaches will affect niche, anonymous communities composed of individuals who are already predisposed to conspiratorial thinking or ``corrupted epistemology", as coined by Sunstein \cite{sunstein2018republic}, reducing wide exposure to such content is likely a social good.
{Furthermore, artificially reducing perceived demand for such content may drive a pro-social incentive for content creators to produce higher quality or less harmful content.}
As concerns about YouTube and its role in misinforming and radicalizing individuals increases, it is good to see the platform take steps to address these issues that appear to have an effect rather than being simply lip service. 

From a general perspective, the evidence outlined herein suggests removing potentially harmful content from recommendation could strike an effective balance between allowing anti-social or subversive views to remain available without actively promoting and propagating such views. 
That is, individuals have the right to share their views but do not have the right to the platform's assistance in monetizing those views.
Other platforms employ similar strategies to suppress undesirable content, as explored by Reddit's quarantining subreddits and WhatsApp's preventing individuals from sharing links more than five times. 
Facebook likewise suppresses (though does not remove) content fact-checkers have identified as untrue, and Twitter  both tags and suppresses anti-social content that violates its community guidelines.
More draconian approaches could disable native retweeting of specific content or disable native sharing for anti-social content; critically, individuals could still \emph{share} this content manually, but adding an extra step to the propagation of such content could have a wide impact.

Without further research into the unintended consequences of these interventions, however, the public should be concerned about what kinds of content get propagated instead. 
The growth we see in content from conspiracy-oriented channels may be one such unintended consequence, with borderline  content that was not classified as harmful receiving \emph{more} recommendations in place of the removed content.
Despite this possibility, advancing this research is hindered by platform opacity, as we get little insight into what borderline content survives YouTube’s classification pipeline, especially if such content is still potentially misinforming; i.e., where is the line between harmful and unharmful content, and does content close to this line end up spreading even more?
Applying such automated classification pipelines without careful and transparent consideration repeats the kinds of choices that brought online platforms to this point in the first place.

\section*{Data Availability}

\hl{YouTube video classifications, time series data, and the analysis for this work is available online in a GitHub repository:} \url{https://github.com/cbuntain/youtube.recsys.xplatform/}

\bibliographystyle{ACM-Reference-Format}  
\bibliography{references}


\begin{thebibliography}{40}


\ifx \showCODEN    \undefined \def \showCODEN     #1{\unskip}     \fi
\ifx \showDOI      \undefined \def \showDOI       #1{#1}\fi
\ifx \showISBNx    \undefined \def \showISBNx     #1{\unskip}     \fi
\ifx \showISBNxiii \undefined \def \showISBNxiii  #1{\unskip}     \fi
\ifx \showISSN     \undefined \def \showISSN      #1{\unskip}     \fi
\ifx \showLCCN     \undefined \def \showLCCN      #1{\unskip}     \fi
\ifx \shownote     \undefined \def \shownote      #1{#1}          \fi
\ifx \showarticletitle \undefined \def \showarticletitle #1{#1}   \fi
\ifx \showURL      \undefined \def \showURL       {\relax}        \fi
\providecommand\bibfield[2]{#2}
\providecommand\bibinfo[2]{#2}
\providecommand\natexlab[1]{#1}
\providecommand\showeprint[2][]{arXiv:#2}

\bibitem[\protect\citeauthoryear{Allcott and Gentzkow}{Allcott and
  Gentzkow}{2017}]%
        {10.1257/jep.31.2.211}
\bibfield{author}{\bibinfo{person}{Hunt Allcott} {and} \bibinfo{person}{Matthew
  Gentzkow}.} \bibinfo{year}{2017}\natexlab{}.
\newblock \showarticletitle{{Social Media and Fake News in the 2016 Election}}.
\newblock \bibinfo{journal}{\emph{Journal of Economic Perspectives}}
  \bibinfo{volume}{31}, \bibinfo{number}{2} (\bibinfo{date}{May}
  \bibinfo{year}{2017}), \bibinfo{pages}{211--236}.
\newblock
\urldef\tempurl%
\url{https://doi.org/10.1257/jep.31.2.211}
\showDOI{\tempurl}


\bibitem[\protect\citeauthoryear{Baumgartner, Zannettou, Keegan, Squire, and
  Blackburn}{Baumgartner et~al\mbox{.}}{2020}]%
        {baumgartner2020pushshift}
\bibfield{author}{\bibinfo{person}{Jason Baumgartner}, \bibinfo{person}{Savvas
  Zannettou}, \bibinfo{person}{Brian Keegan}, \bibinfo{person}{Megan Squire},
  {and} \bibinfo{person}{Jeremy Blackburn}.} \bibinfo{year}{2020}\natexlab{}.
\newblock \showarticletitle{{The Pushshift Reddit Dataset}}. In
  \bibinfo{booktitle}{\emph{arXiv preprint}}.
\newblock
\showeprint[arxiv]{cs.SI/2001.08435}
\urldef\tempurl%
\url{http://arxiv.org/abs/2001.08435}
\showURL{%
\tempurl}


\bibitem[\protect\citeauthoryear{Bernal, Cummins, and Gasparrini}{Bernal
  et~al\mbox{.}}{2017}]%
        {Bernal2017}
\bibfield{author}{\bibinfo{person}{James~Lopez Bernal}, \bibinfo{person}{Steven
  Cummins}, {and} \bibinfo{person}{Antonio Gasparrini}.}
  \bibinfo{year}{2017}\natexlab{}.
\newblock \showarticletitle{{Interrupted Time Series Regression for the
  Evaluation of Public Health Interventions: A Tutorial}}.
\newblock \bibinfo{journal}{\emph{International Journal of Epidemiology}}
  \bibinfo{volume}{46}, \bibinfo{number}{1} (\bibinfo{year}{2017}),
  \bibinfo{pages}{348--355}.
\newblock
\showISSN{14643685}
\urldef\tempurl%
\url{https://doi.org/10.1093/ije/dyw098}
\showDOI{\tempurl}


\bibitem[\protect\citeauthoryear{Bernal, Cummins, and Gasparrini}{Bernal
  et~al\mbox{.}}{2018}]%
        {Bernal2018}
\bibfield{author}{\bibinfo{person}{James~Lopez Bernal}, \bibinfo{person}{Steven
  Cummins}, {and} \bibinfo{person}{Antonio Gasparrini}.}
  \bibinfo{year}{2018}\natexlab{}.
\newblock \showarticletitle{{The Use of Controls In Interrupted Time Series
  Studies of Public Health Interventions}}.
\newblock \bibinfo{journal}{\emph{International Journal of Epidemiology}}
  \bibinfo{volume}{47}, \bibinfo{number}{6} (\bibinfo{year}{2018}),
  \bibinfo{pages}{2082--2093}.
\newblock
\showISSN{14643685}
\urldef\tempurl%
\url{https://doi.org/10.1093/ije/dyy135}
\showDOI{\tempurl}


\bibitem[\protect\citeauthoryear{Bursztyn, Rao, Roth, and
  Yanagizawa-Drott}{Bursztyn et~al\mbox{.}}{2020}]%
        {Bursztyn2020}
\bibfield{author}{\bibinfo{person}{Leonardo Bursztyn}, \bibinfo{person}{Aakaash
  Rao}, \bibinfo{person}{Christopher Roth}, {and} \bibinfo{person}{David
  Yanagizawa-Drott}.} \bibinfo{year}{2020}\natexlab{}.
\newblock \showarticletitle{{Misinformation During a Pandemic}}.
\newblock \bibinfo{journal}{\emph{SSRN Electronic Journal}}
  (\bibinfo{year}{2020}).
\newblock
\urldef\tempurl%
\url{https://doi.org/10.2139/ssrn.3580487}
\showDOI{\tempurl}


\bibitem[\protect\citeauthoryear{Chaintreau, Ramachandran, and Wang}{Chaintreau
  et~al\mbox{.}}{2016}]%
        {Chaintreau2016}
\bibfield{author}{\bibinfo{person}{Augustin Chaintreau}, \bibinfo{person}{Arthi
  Ramachandran}, {and} \bibinfo{person}{Lucy~X. Wang}.}
  \bibinfo{year}{2016}\natexlab{}.
\newblock \showarticletitle{{Measuring Click and Share Dynamics on Social
  Media: A Reproducible and Validated Approach}}.
\newblock \bibinfo{journal}{\emph{AAAI Workshop - Technical Report}}
  \bibinfo{volume}{WS-16-16 - WS-16-20} (\bibinfo{year}{2016}),
  \bibinfo{pages}{108--113}.
\newblock
\showISBNx{9781577357681}


\bibitem[\protect\citeauthoryear{Chancellor, Pater, Clear, Gilbert, and {De
  Choudhury}}{Chancellor et~al\mbox{.}}{2016}]%
        {Chancellor:2016:TIC:2818048.2819963}
\bibfield{author}{\bibinfo{person}{Stevie Chancellor},
  \bibinfo{person}{Jessica~Annette Pater}, \bibinfo{person}{Trustin Clear},
  \bibinfo{person}{Eric Gilbert}, {and} \bibinfo{person}{Munmun {De
  Choudhury}}.} \bibinfo{year}{2016}\natexlab{}.
\newblock \showarticletitle{{{\#}Thyghgapp: Instagram Content Moderation and
  Lexical Variation in Pro-Eating Disorder Communities}}. In
  \bibinfo{booktitle}{\emph{Proceedings of the 19th ACM Conference on
  Computer-Supported Cooperative Work {\&} Social Computing}}
  \emph{(\bibinfo{series}{CSCW '16})}. \bibinfo{publisher}{ACM},
  \bibinfo{address}{New York, NY, USA}, \bibinfo{pages}{1201--1213}.
\newblock
\showISBNx{978-1-4503-3592-8}
\urldef\tempurl%
\url{https://doi.org/10.1145/2818048.2819963}
\showDOI{\tempurl}


\bibitem[\protect\citeauthoryear{Chandrasekharan, Pavalanathan, Srinivasan,
  Glynn, Eisenstein, and Gilbert}{Chandrasekharan et~al\mbox{.}}{2017}]%
        {Chandrasekharan:2017:YCS:3171581.3134666}
\bibfield{author}{\bibinfo{person}{Eshwar Chandrasekharan},
  \bibinfo{person}{Umashanthi Pavalanathan}, \bibinfo{person}{Anirudh
  Srinivasan}, \bibinfo{person}{Adam Glynn}, \bibinfo{person}{Jacob
  Eisenstein}, {and} \bibinfo{person}{Eric Gilbert}.}
  \bibinfo{year}{2017}\natexlab{}.
\newblock \showarticletitle{{You Can'T Stay Here: The Efficacy of Reddit's 2015
  Ban Examined Through Hate Speech}}.
\newblock \bibinfo{journal}{\emph{Proc. ACM Hum.-Comput. Interact.}}
  \bibinfo{volume}{1}, \bibinfo{number}{CSCW} (\bibinfo{date}{Dec}
  \bibinfo{year}{2017}), \bibinfo{pages}{31:1----31:22}.
\newblock
\showISSN{2573-0142}
\urldef\tempurl%
\url{https://doi.org/10.1145/3134666}
\showDOI{\tempurl}


\bibitem[\protect\citeauthoryear{Cinelli, Quattrociocchi, Galeazzi, Valensise,
  Brugnoli, Schmidt, Zola, Zollo, and Scala}{Cinelli et~al\mbox{.}}{2020}]%
        {Cinelli2020}
\bibfield{author}{\bibinfo{person}{Matteo Cinelli}, \bibinfo{person}{Walter
  Quattrociocchi}, \bibinfo{person}{Alessandro Galeazzi},
  \bibinfo{person}{Carlo~Michele Valensise}, \bibinfo{person}{Emanuele
  Brugnoli}, \bibinfo{person}{Ana~Lucia Schmidt}, \bibinfo{person}{Paola Zola},
  \bibinfo{person}{Fabiana Zollo}, {and} \bibinfo{person}{Antonio Scala}.}
  \bibinfo{year}{2020}\natexlab{}.
\newblock \showarticletitle{{The COVID-19 Social Media Infodemic}}.
\newblock \bibinfo{journal}{\emph{arXiv preprint}} (\bibinfo{year}{2020}),
  \bibinfo{pages}{1--18}.
\newblock
\showeprint[arxiv]{2003.05004}
\urldef\tempurl%
\url{http://arxiv.org/abs/2003.05004}
\showURL{%
\tempurl}


\bibitem[\protect\citeauthoryear{Facebook}{Facebook}{2018}]%
        {Facebook2018}
\bibfield{author}{\bibinfo{person}{Facebook}.} \bibinfo{year}{2018}\natexlab{}.
\newblock \bibinfo{title}{{Hard Questions: What's Facebook's Strategy for
  Stopping False News?}}
\newblock
\newblock
\urldef\tempurl%
\url{https://about.fb.com/news/2018/05/hard-questions-false-news/}
\showURL{%
\tempurl}


\bibitem[\protect\citeauthoryear{Faddoul, Chaslot, and Farid}{Faddoul
  et~al\mbox{.}}{2020}]%
        {Faddoul2020}
\bibfield{author}{\bibinfo{person}{Marc Faddoul}, \bibinfo{person}{Guillaume
  Chaslot}, {and} \bibinfo{person}{Hany Farid}.}
  \bibinfo{year}{2020}\natexlab{}.
\newblock \showarticletitle{{A Longitudinal Analysis of YouTube's Promotion of
  Conspiracy Videos}}. In \bibinfo{booktitle}{\emph{arXiv preprint}}.
  \bibinfo{pages}{1--8}.
\newblock
\showeprint[arxiv]{2003.03318}
\urldef\tempurl%
\url{http://arxiv.org/abs/2003.03318}
\showURL{%
\tempurl}


\bibitem[\protect\citeauthoryear{Greenspan}{Greenspan}{2019}]%
        {Greenspan2019}
\bibfield{author}{\bibinfo{person}{Rachel Greenspan}.}
  \bibinfo{year}{2019}\natexlab{}.
\newblock \showarticletitle{{WhatsApp Fights Fake News With New Message
  Restrictions | Time}}.
\newblock \bibinfo{journal}{\emph{Time}} (\bibinfo{date}{Jan}
  \bibinfo{year}{2019}), \bibinfo{pages}{21}.
\newblock
\urldef\tempurl%
\url{https://time.com/5508630/whatsapp-message-restrictions/}
\showURL{%
\tempurl}


\bibitem[\protect\citeauthoryear{Grinberg, Joseph, Friedland, Swire-Thompson,
  and Lazer}{Grinberg et~al\mbox{.}}{2019}]%
        {Grinberg2019}
\bibfield{author}{\bibinfo{person}{Nir Grinberg}, \bibinfo{person}{Kenneth
  Joseph}, \bibinfo{person}{Lisa Friedland}, \bibinfo{person}{Briony
  Swire-Thompson}, {and} \bibinfo{person}{David Lazer}.}
  \bibinfo{year}{2019}\natexlab{}.
\newblock \showarticletitle{{Political Science: Fake News on Twitter During the
  2016 U.S. presidential election}}.
\newblock \bibinfo{journal}{\emph{Science}} \bibinfo{volume}{363},
  \bibinfo{number}{6425} (\bibinfo{year}{2019}), \bibinfo{pages}{374--378}.
\newblock
\showISSN{10959203}
\urldef\tempurl%
\url{https://doi.org/10.1126/science.aau2706}
\showDOI{\tempurl}


\bibitem[\protect\citeauthoryear{Guess, Nyhan, and Reifler}{Guess
  et~al\mbox{.}}{2018}]%
        {Guess2018}
\bibfield{author}{\bibinfo{person}{Andrew Guess}, \bibinfo{person}{Brendan
  Nyhan}, {and} \bibinfo{person}{Jason Reifler}.}
  \bibinfo{year}{2018}\natexlab{}.
\newblock \bibinfo{booktitle}{\emph{{Selective Exposure to Misinformation:
  Evidence from the Consumption of Fake News During the 2016 U.S. Presidential
  Campaign}}}.
\newblock \bibinfo{type}{{T}echnical {R}eport} 682758.
  \bibinfo{institution}{European Research Council}. \bibinfo{pages}{1--34}
  pages.
\newblock
\urldef\tempurl%
\url{https://www.dartmouth.edu/{~}nyhan/fake-news-2016.pdf}
\showURL{%
\tempurl}


\bibitem[\protect\citeauthoryear{Jansen and Martin}{Jansen and Martin}{2015}]%
        {Jansen2015}
\bibfield{author}{\bibinfo{person}{Sue~Curry Jansen} {and}
  \bibinfo{person}{Brian Martin}.} \bibinfo{year}{2015}\natexlab{}.
\newblock \showarticletitle{{The Streisand Effect and Censorship Backfire}}.
\newblock \bibinfo{journal}{\emph{International Journal of Communication}}
  \bibinfo{volume}{9}, \bibinfo{number}{1} (\bibinfo{year}{2015}),
  \bibinfo{pages}{656--671}.
\newblock
\showISSN{19328036}


\bibitem[\protect\citeauthoryear{Jurkowitz, Mitchell, Shearer, and
  Walker}{Jurkowitz et~al\mbox{.}}{2020}]%
        {Jurkowitz2020}
\bibfield{author}{\bibinfo{person}{Mark Jurkowitz}, \bibinfo{person}{Amy
  Mitchell}, \bibinfo{person}{Elisa Shearer}, {and} \bibinfo{person}{Mason
  Walker}.} \bibinfo{year}{2020}\natexlab{}.
\newblock \showarticletitle{{U.S. Media Polarization and the 2020 Election: A
  Nation Divided}}.
\newblock \bibinfo{journal}{\emph{Pew Research Centre}} (\bibinfo{year}{2020}),
  \bibinfo{pages}{67}.
\newblock
\urldef\tempurl%
\url{https://www.journalism.org/2020/01/24/americans-are-divided-by-party-in-the-sources-they-turn-to-for-political-news/}
\showURL{%
\tempurl}


\bibitem[\protect\citeauthoryear{Ledwich and Zaitsev}{Ledwich and
  Zaitsev}{2020}]%
        {Ledwich2020}
\bibfield{author}{\bibinfo{person}{Mark Ledwich} {and} \bibinfo{person}{Anna
  Zaitsev}.} \bibinfo{year}{2020}\natexlab{}.
\newblock \showarticletitle{{Algorithmic Extremism: Examining YouTube's Rabbit
  Hole of Radicalization}}.
\newblock \bibinfo{journal}{\emph{First Monday}} (\bibinfo{year}{2020}).
\newblock
\showISSN{1396-0466}
\urldef\tempurl%
\url{https://doi.org/10.5210/fm.v25i3.10419}
\showDOI{\tempurl}
\showeprint[arxiv]{1912.11211}


\bibitem[\protect\citeauthoryear{Leskovec, Rajaraman, and Ullman}{Leskovec
  et~al\mbox{.}}{2014}]%
        {leskovec_rajaraman_ullman_2014}
\bibfield{author}{\bibinfo{person}{Jure Leskovec}, \bibinfo{person}{Anand
  Rajaraman}, {and} \bibinfo{person}{Jeffrey~David Ullman}.}
  \bibinfo{year}{2014}\natexlab{}.
\newblock \bibinfo{booktitle}{\emph{{Mining of Massive Datasets}}
  (\bibinfo{edition}{2} ed.)}.
\newblock \bibinfo{publisher}{Cambridge University Press}.
\newblock
\urldef\tempurl%
\url{https://doi.org/10.1017/CBO9781139924801}
\showDOI{\tempurl}


\bibitem[\protect\citeauthoryear{Lewis}{Lewis}{2018}]%
        {Lewis2018}
\bibfield{author}{\bibinfo{person}{Rebecca Lewis}.}
  \bibinfo{year}{2018}\natexlab{}.
\newblock \showarticletitle{{Alternative Influence: Broadcasting the
  Reactionary Right on YouTube}}.
\newblock \bibinfo{journal}{\emph{Data {\&} Society}} (\bibinfo{year}{2018}).
\newblock
\urldef\tempurl%
\url{https://datasociety.net/research/media-manipulation.}
\showURL{%
\tempurl}


\bibitem[\protect\citeauthoryear{Mariconti, Suarez-Tangil, Blackburn, {De
  Cristofaro}, Kourtellis, Leontiadis, Serrano, and Stringhini}{Mariconti
  et~al\mbox{.}}{2019}]%
        {Mariconti2019}
\bibfield{author}{\bibinfo{person}{Enrico Mariconti},
  \bibinfo{person}{Guillermo Suarez-Tangil}, \bibinfo{person}{Jeremy
  Blackburn}, \bibinfo{person}{Emiliano {De Cristofaro}},
  \bibinfo{person}{Nicolas Kourtellis}, \bibinfo{person}{Ilias Leontiadis},
  \bibinfo{person}{Jordi~Luque Serrano}, {and} \bibinfo{person}{Gianluca
  Stringhini}.} \bibinfo{year}{2019}\natexlab{}.
\newblock \showarticletitle{{“You Know What To Do”: Proactive Detection of
  YouTube Videos Targeted By Coordinated Hate Attacks}}.
\newblock \bibinfo{journal}{\emph{Proceedings of the ACM on Human-Computer
  Interaction}} \bibinfo{volume}{3}, \bibinfo{number}{CSCW}
  (\bibinfo{year}{2019}).
\newblock
\showISSN{25730142}
\urldef\tempurl%
\url{https://doi.org/10.1145/3359309}
\showDOI{\tempurl}
\showeprint[arxiv]{1805.08168}


\bibitem[\protect\citeauthoryear{Morstatter, Pfeffer, Liu, and
  Carley}{Morstatter et~al\mbox{.}}{2013}]%
        {Morstatter2013}
\bibfield{author}{\bibinfo{person}{Fred Morstatter}, \bibinfo{person}{J
  Pfeffer}, \bibinfo{person}{H Liu}, {and} \bibinfo{person}{Km Carley}.}
  \bibinfo{year}{2013}\natexlab{}.
\newblock \showarticletitle{{Is the Sample Good Enough? Comparing Data from
  Twitter's Streaming API with Twitter's Firehose}}.
\newblock \bibinfo{journal}{\emph{Proceedings of ICWSM}}
  (\bibinfo{year}{2013}), \bibinfo{pages}{400--408}.
\newblock
\showISBNx{9783319055787}
\showISSN{16113349}
\urldef\tempurl%
\url{https://doi.org/10.1007/978-3-319-05579-4_10}
\showDOI{\tempurl}
\showeprint[arxiv]{arXiv:1306.5204v1}


\bibitem[\protect\citeauthoryear{Munger}{Munger}{2019}]%
        {Munger_2019}
\bibfield{author}{\bibinfo{person}{Kevin Munger}.}
  \bibinfo{year}{2019}\natexlab{}.
\newblock \bibinfo{title}{{YouTube Politics}}.
\newblock
\newblock
\urldef\tempurl%
\url{osf.io/4wk63}
\showURL{%
\tempurl}


\bibitem[\protect\citeauthoryear{Munger and Phillips}{Munger and
  Phillips}{2019}]%
        {Munger2019}
\bibfield{author}{\bibinfo{person}{Kevin Munger} {and} \bibinfo{person}{Joseph
  Phillips}.} \bibinfo{year}{2019}\natexlab{}.
\newblock \showarticletitle{{A Supply and Demand Framework for YouTube
  Politics}}.
\newblock  (\bibinfo{year}{2019}).
\newblock


\bibitem[\protect\citeauthoryear{{Myers West}}{{Myers West}}{2018}]%
        {MyersWest2018}
\bibfield{author}{\bibinfo{person}{Sarah {Myers West}}.}
  \bibinfo{year}{2018}\natexlab{}.
\newblock \showarticletitle{{Censored, Suspended, Shadowbanned: User
  Interpretations of Content Moderation on Social Media Platforms}}.
\newblock \bibinfo{journal}{\emph{New Media and Society}} \bibinfo{volume}{20},
  \bibinfo{number}{11} (\bibinfo{year}{2018}), \bibinfo{pages}{4366--4383}.
\newblock
\showISSN{14617315}
\urldef\tempurl%
\url{https://doi.org/10.1177/1461444818773059}
\showDOI{\tempurl}


\bibitem[\protect\citeauthoryear{Neil and Campbell}{Neil and Campbell}{2020}]%
        {Neil2020}
\bibfield{author}{\bibinfo{person}{Stuart Neil} {and} \bibinfo{person}{Edward~M
  Campbell}.} \bibinfo{year}{2020}\natexlab{}.
\newblock \showarticletitle{{Fake Science: Xmrv, Covid-19 and the Toxic Legacy
  of Dr Judy Mikovits}}.
\newblock \bibinfo{journal}{\emph{AIDS Research and Human Retroviruses}}
  \bibinfo{volume}{00}, \bibinfo{number}{00} (\bibinfo{year}{2020}),
  \bibinfo{pages}{1--5}.
\newblock
\showISSN{0889-2229}
\urldef\tempurl%
\url{https://doi.org/10.1089/aid.2020.0095}
\showDOI{\tempurl}


\bibitem[\protect\citeauthoryear{Newell, Jurgens, Saleem, Vala, Sassine,
  Armstrong, and Ruths}{Newell et~al\mbox{.}}{2016}]%
        {Newell2016}
\bibfield{author}{\bibinfo{person}{Edward Newell}, \bibinfo{person}{David
  Jurgens}, \bibinfo{person}{Haji~Mohammad Saleem}, \bibinfo{person}{Hardik
  Vala}, \bibinfo{person}{Jad Sassine}, \bibinfo{person}{Caitrin Armstrong},
  {and} \bibinfo{person}{Derek Ruths}.} \bibinfo{year}{2016}\natexlab{}.
\newblock \showarticletitle{{User Migration in Online Social Networks: A Case
  Study on Reddit During a Period of Community Unrest}}.
\newblock \bibinfo{journal}{\emph{Proceedings of the 10th International
  Conference on Web and Social Media, ICWSM 2016}} \bibinfo{number}{Icwsm}
  (\bibinfo{year}{2016}), \bibinfo{pages}{279--288}.
\newblock
\showISBNx{9781577357582}


\bibitem[\protect\citeauthoryear{Ribeiro, Ottoni, West, Almeida, and
  Meira}{Ribeiro et~al\mbox{.}}{2019}]%
        {Ribeiro2019}
\bibfield{author}{\bibinfo{person}{Manoel~Horta Ribeiro},
  \bibinfo{person}{Raphael Ottoni}, \bibinfo{person}{Robert West},
  \bibinfo{person}{Virg{\'{i}}lio A.~F. Almeida}, {and} \bibinfo{person}{Wagner
  Meira}.} \bibinfo{year}{2019}\natexlab{}.
\newblock \showarticletitle{{Auditing Radicalization Pathways on YouTube}}.
\newblock \bibinfo{journal}{\emph{arXiv preprint arXiv:1908.08313}}
  (\bibinfo{year}{2019}).
\newblock
\showeprint[arxiv]{1908.08313}
\urldef\tempurl%
\url{http://arxiv.org/abs/1908.08313}
\showURL{%
\tempurl}


\bibitem[\protect\citeauthoryear{Smith and Anderson}{Smith and
  Anderson}{2018}]%
        {Smith2018}
\bibfield{author}{\bibinfo{person}{Aaron Smith} {and} \bibinfo{person}{Monica
  Anderson}.} \bibinfo{year}{2018}\natexlab{}.
\newblock \bibinfo{booktitle}{\emph{{Social Media Use in 2018}}}.
\newblock \bibinfo{type}{{T}echnical {R}eport} March. \bibinfo{institution}{Pew
  Research Center}. \bibinfo{pages}{4} pages.
\newblock


\bibitem[\protect\citeauthoryear{Starbird}{Starbird}{2017}]%
        {Starbird2017}
\bibfield{author}{\bibinfo{person}{Kate Starbird}.}
  \bibinfo{year}{2017}\natexlab{}.
\newblock \showarticletitle{{Examining the Alternative Media Ecosystem Through
  the Production of Alternative Narratives of Mass Shooting Events on
  Twitter}}.
\newblock \bibinfo{journal}{\emph{International AAAI Conference on Web and
  Social Media (ICWSM)}} (\bibinfo{year}{2017}).
\newblock
\urldef\tempurl%
\url{http://faculty.washington.edu/kstarbi/Alt{\_}Narratives{\_}ICWSM17-CameraReady.pdf}
\showURL{%
\tempurl}


\bibitem[\protect\citeauthoryear{Sunstein}{Sunstein}{2018}]%
        {sunstein2018republic}
\bibfield{author}{\bibinfo{person}{C~R Sunstein}.}
  \bibinfo{year}{2018}\natexlab{}.
\newblock \bibinfo{booktitle}{\emph{{{\#}Republic: Divided Democracy in the Age
  of Social Media}}}.
\newblock \bibinfo{publisher}{Princeton University Press}.
\newblock
\showISBNx{9780691180908}
\urldef\tempurl%
\url{https://books.google.com/books?id=nVBLDwAAQBAJ}
\showURL{%
\tempurl}


\bibitem[\protect\citeauthoryear{Suzor}{Suzor}{2019}]%
        {Suzor2019}
\bibfield{author}{\bibinfo{person}{Nicolas Suzor}.}
  \bibinfo{year}{2019}\natexlab{}.
\newblock \bibinfo{booktitle}{\emph{{YouTube Stops Recommending Alt-Right
  Videos - Digital Social Contract}}}.
\newblock \bibinfo{type}{{T}echnical {R}eport}.
  \bibinfo{institution}{Queensland University of Technology}.
\newblock
\urldef\tempurl%
\url{https://digitalsocialcontract.net/youtube-stops-recommending-alt-right-videos-6523ed6af60f}
\showURL{%
\tempurl}


\bibitem[\protect\citeauthoryear{Trujillo, Gruppi, Buntain, and Horne}{Trujillo
  et~al\mbox{.}}{2020}]%
        {10.1145/3372923.3404833}
\bibfield{author}{\bibinfo{person}{Milo Trujillo}, \bibinfo{person}{Mauricio
  Gruppi}, \bibinfo{person}{Cody Buntain}, {and} \bibinfo{person}{Benjamin~D
  Horne}.} \bibinfo{year}{2020}\natexlab{}.
\newblock \showarticletitle{{What is BitChute? Characterizing the "Free Speech"
  Alternative to YouTube}}. In \bibinfo{booktitle}{\emph{Proceedings of the
  31st ACM Conference on Hypertext and Social Media}}
  \emph{(\bibinfo{series}{HT '20})}. \bibinfo{publisher}{Association for
  Computing Machinery}, \bibinfo{address}{New York, NY, USA},
  \bibinfo{pages}{139--140}.
\newblock
\showISBNx{9781450370981}
\urldef\tempurl%
\url{https://doi.org/10.1145/3372923.3404833}
\showDOI{\tempurl}


\bibitem[\protect\citeauthoryear{Tufekci}{Tufekci}{2018}]%
        {Tufekci2018}
\bibfield{author}{\bibinfo{person}{Zeynep Tufekci}.}
  \bibinfo{year}{2018}\natexlab{}.
\newblock \bibinfo{title}{{Opinion | YouTube, the Great Radicalizer}}.
\newblock
\newblock
\urldef\tempurl%
\url{https://www.nytimes.com/2018/03/10/opinion/sunday/youtube-politics-radical.html}
\showURL{%
\tempurl}


\bibitem[\protect\citeauthoryear{Vigdor and Chokshi}{Vigdor and
  Chokshi}{2019}]%
        {Vigdor2019}
\bibfield{author}{\bibinfo{person}{Neil Vigdor} {and} \bibinfo{person}{Niraj
  Chokshi}.} \bibinfo{year}{2019}\natexlab{}.
\newblock \bibinfo{title}{{Reddit Restricts Pro-Trump Forum Because of Threats
  - The New York Times}}.
\newblock
\newblock
\urldef\tempurl%
\url{https://www.nytimes.com/2019/06/26/us/politics/reddit-donald-trump-quarantined.html}
\showURL{%
\tempurl}


\bibitem[\protect\citeauthoryear{Wakabayashi}{Wakabayashi}{2019}]%
        {Wakabayashi2019}
\bibfield{author}{\bibinfo{person}{Daisuke Wakabayashi}.}
  \bibinfo{year}{2019}\natexlab{}.
\newblock \bibinfo{title}{{YouTube Moves to Make Conspiracy Videos Harder to
  Find}}.
\newblock
\newblock
\urldef\tempurl%
\url{https://www.nytimes.com/2019/01/25/technology/youtube-conspiracy-theory-videos.html}
\showURL{%
\tempurl}


\bibitem[\protect\citeauthoryear{Yin}{Yin}{2018}]%
        {leon_yin_2018_1345144}
\bibfield{author}{\bibinfo{person}{Leon Yin}.} \bibinfo{year}{2018}\natexlab{}.
\newblock \bibinfo{title}{{SMAPPNYU/urlExpander: Initial release}}.
\newblock
\newblock
\urldef\tempurl%
\url{https://doi.org/10.5281/zenodo.1345144}
\showDOI{\tempurl}


\bibitem[\protect\citeauthoryear{Yin and Brown}{Yin and Brown}{2018}]%
        {leon_yin_2018_1414418}
\bibfield{author}{\bibinfo{person}{Leon Yin} {and} \bibinfo{person}{Megan
  Brown}.} \bibinfo{year}{2018}\natexlab{}.
\newblock \bibinfo{title}{{SMAPPNYU/youtube-data-api}}.
\newblock
\newblock
\urldef\tempurl%
\url{https://doi.org/10.5281/zenodo.1414418}
\showDOI{\tempurl}


\bibitem[\protect\citeauthoryear{{YouTube Team}}{{YouTube Team}}{2019a}]%
        {YouTubeTeam2019}
\bibfield{author}{\bibinfo{person}{{YouTube Team}}.}
  \bibinfo{year}{2019}\natexlab{a}.
\newblock \bibinfo{title}{{Continuing Our Work to Improve Recommendations on
  YouTube}}.
\newblock
\newblock
\urldef\tempurl%
\url{https://youtube.googleblog.com/2019/01/continuing-our-work-to-improve.html}
\showURL{%
\tempurl}


\bibitem[\protect\citeauthoryear{{YouTube Team}}{{YouTube Team}}{2019b}]%
        {YouTubeTeam}
\bibfield{author}{\bibinfo{person}{{YouTube Team}}.}
  \bibinfo{year}{2019}\natexlab{b}.
\newblock \bibinfo{title}{{Our ongoing work to tackle hate}}.
\newblock
\newblock
\urldef\tempurl%
\url{https://youtube.googleblog.com/2019/06/our-ongoing-work-to-tackle-hate.html}
\showURL{%
\tempurl}


\bibitem[\protect\citeauthoryear{Zannettou, Caulfield, {De Cristofaro},
  Kourtellis, Leontiadis, Sirivianos, Stringhini, Blackburn, Kourtelris,
  Leontiadis, Sirivianos, Stringhini, and Blackburn}{Zannettou
  et~al\mbox{.}}{2017}]%
        {Zannettou:2017:WCU:3131365.3131390}
\bibfield{author}{\bibinfo{person}{Savvas Zannettou}, \bibinfo{person}{Tristan
  Caulfield}, \bibinfo{person}{Emiliano {De Cristofaro}},
  \bibinfo{person}{Nicolas Kourtellis}, \bibinfo{person}{Ilias Leontiadis},
  \bibinfo{person}{Michael Sirivianos}, \bibinfo{person}{Gianluca Stringhini},
  \bibinfo{person}{Jeremy Blackburn}, \bibinfo{person}{Nicolas Kourtelris},
  \bibinfo{person}{Ilias Leontiadis}, \bibinfo{person}{Michael Sirivianos},
  \bibinfo{person}{Gianluca Stringhini}, {and} \bibinfo{person}{Jeremy
  Blackburn}.} \bibinfo{year}{2017}\natexlab{}.
\newblock \showarticletitle{{The Web Centipede: Understanding How Web
  Communities Influence Each Other Through the Lens of Mainstream and
  Alternative News Sources}}. In \bibinfo{booktitle}{\emph{Proceedings of the
  2017 Internet Measurement Conference}} \emph{(\bibinfo{series}{IMC '17})}.
  \bibinfo{publisher}{ACM}, \bibinfo{address}{New York, NY, USA},
  \bibinfo{pages}{405--417}.
\newblock
\showISBNx{978-1-4503-5118-8}
\showISSN{08955638}
\urldef\tempurl%
\url{https://doi.org/10.1145/3131365.3131390}
\showDOI{\tempurl}
\showeprint[arxiv]{1705.06947}


\end{thebibliography}

\end{document}